\documentclass[12pt]{article}
\usepackage{graphicx}
\usepackage{mathptmx,mathrsfs}
\usepackage{amsfonts,amsmath,amssymb,amsthm}
\usepackage{indentfirst}
\usepackage{cite,hyperref}
\usepackage{enumitem}
\usepackage{color}

\numberwithin{equation}{section}

\theoremstyle{definition}

\newcommand{\email}[1]{\footnote{E-mail: \href{mailto:#1}{#1}}}
\begin{document}

\title{Normalizability analysis of the generalized quantum electrodynamics
from the causal point of view}
\author{R.~Bufalo$^{a}$ \email{%
rodrigo.bufalo@dfi.ufla.br}~, B.M.~Pimentel$^{b}$ \email{%
pimentel@ift.unesp.br}~, and D.E.~Soto$^{c}$ \email{%
dsotob@uni.edu.pe}~ \\
\textit{$^{a}$ \small Departamento de F\'isica, Universidade Federal de Lavras,}\\
\textit{ \small Caixa Postal 3037, 37200-000 Lavras, MG, Brazil}\\
\textit{ $^{b}$ \small Instituto de F\'{\i}sica Te\'orica, Universidade Estadual Paulista} \\
\textit{\small Rua Dr. Bento Teobaldo Ferraz 271, Bloco II Barra Funda, S\~ao Paulo -- SP, 01140-070, Brazil}\\
\textit{$^{c}$ \small Facultad de Ciencias, Universidad Nacional de Ingenier\'ia UNI}\\
\textit{ \small Avenida T\'upac Amaru S/N apartado 31139 Lima, Per\'u}\\
}

\maketitle
\date{}

\begin{abstract}
The causal perturbation theory is an axiomatic perturbative theory of the
S-matrix. This formalism has as its essence the following axioms: causality,
Lorentz invariance and asymptotic conditions. Any other property must be showed
via the inductive method order-by-order and, of course, it depends on
the particular physical model. In this work we shall study the normalizability
of the generalized quantum electrodynamics in the framework of the causal
approach. Furthermore, we analyse the implication of the gauge invariance
onto the model and obtain the respective Ward-Takahashi-Fradkin identities.
\end{abstract}
\begin{flushleft}
\end{flushleft}

\newpage
\tableofcontents


\section{Introduction}

\label{sec:1}

In the usual approach, the main idea behind renormalization program is that
ultraviolet (UV) divergences of a field theory are to be absorbed by an
appropriated renormalization of the parameters of the theory \cite{ref29,ref30,ref76}.
\footnote{Infrared (IR) divergences may as well appear and these come from the existence of long-range forces in the theory.} For perturbative field theories this process is performed order-by-order for each class of graphs.
Besides, it is known that the majority of physical field theories, described by
first-order Lagrangians, are usually plagued with UV and IR divergences.
More importantly, it is possible to know whether the renormalization program is applicable to a particular model; for a perturbative theory
this can be determined by the method of dimensional analysis and
power-counting \cite{ref32}.
Thus, we say that a theory is renormalizable if all the necessary counterterms
are found directly from the original Lagrangian.

It is a well known fact that higher-order derivative (HD) theories \cite{ref1} have,
in light of effective field theory \cite{ref2}, better
renormalizability properties than the conventional ones.
This idea is rather successful in the case of an attempt to quantize gravity, where the
(non-renormalizable) Einstein action is supplied by terms containing higher powers of
curvature leading to a renormalizable \cite{ref4} and asymptotically free theory \cite{ref5}.
Also, we can refer to an impetus in exploring appealing quantum gravitational theories, such as
$f\left( R\right) $-gravity \cite{ref6} and Horava-Lifshitz gravity \cite{ref77}.
However, it was soon recognized that HD theories have a Hamiltonian which is not bounded
from below \cite{ref7} and that the addition of such terms leads to the existence of negative
norm states (or ghosts states) -- induces an indefinite metric in the space of states -- jeopardizing
thus the unitarity \cite{ref8}. Despite the fact that many attempts to overcome these ghost states
have been proposed, no one has been able to give a general method to deal with them \cite{ref9,ref10}.

One of the most simple and interesting contributions to show the effectiveness of
HD terms in field theory were due to Bopp \cite{ref12} and Podolsky and
Schwed \cite{ref13}, who proposed a generalization of the Maxwell
electromagnetic field. \footnote{A non-Abelian version of the Bopp-Podolsky
electrodynamics was studied and deeply analyzed in \cite{ref3,ref38}
as well its interaction with the gravitational field \cite{ref84}.} 
The quantum-particle of this field is called Podolsky photon and the interaction of these quanta with electrons is known as generalized quantum electrodynamics (GQED$_{4}$). 
Moreover, in Ref.~\cite{ref11}, it has been shown that the Podolsky Lagrangian is
the only linear generalization of Maxwell electrodynamics that preserves
invariance under $U\left( 1\right) $. Recently, in Ref.~\cite{ref37} a
procedure was suggested for including interactions in free HD systems
without breaking their stability. Remarkably, they showed that the dynamics
of the GQED$_{4}$ is stable at both classical and quantum level.

It is worth to mention that GQED$_{4}$ have momentous difference with respect to
QED$_{4}$. One of these aspects is the expression of the Born approximation of Bhabha
scattering \cite{ref14}, where the Podolsky mass plays a part of a cut-off term for this process.
Furthermore, in a previous work \cite{ref15} the renormalization program was successfully
applied on GQED$_{4}$; subsequent quantities were computed at one-loop approximation and showed that the self-energy and vertex are UV finite.
Besides, a discussion and evaluation of the Podolsky contribution
to the electron anomalous magnetic moment were addressed in detail.

Nonetheless, there is an alternative and richer approach to analyse the
renormalizability structure of a physical model.
This is the causal perturbation theory, or simply Epstein-Glaser causal theory \cite{ref16}.
In this framework the concept of normalizability is introduced, which is somehow
analogous to the usual renormalizability.
The main difference between them is that in the causal approach every Green's function is
finite order-by-order, thus, the normalization is not a process to subtract divergences, but rather to fix some finite constants.

The causal method is an axiomatic perturbative approach used to systematically compute the elements of the S-matrix \cite{ref16,ref21},
by following closely the Heisenberg program \cite{ref17}.
The causal method takes into account only asymptotic free conditions and some few general properties: causality and Lorentz invariance.
A remarkable advantage of this approach is that any quantity is defined within the framework of distribution theory; hence, all the product of field operators are well-defined at the origin or, equivalently, they are UV finite \cite{ref26}.
This means that no regularization method is necessary to be introduced \cite%
{ref20}.

The causal approach has already been applied in the analysis to prove the
renormalizability of QED$_4$ \cite{ref21} and QED$_3$ \cite{ref19}.
Also in QED$_3$, the Epstein-Glaser theory was used to give a clear solution
of the Pauli-Villar regularization problem \cite{ref18}.
In a similar way, the causal approach helped to prove inductively the normalizability and gauge invariance of the
 scalar quantum electrodynamics (SQED$_{4}$) \cite{ref22}.
Furthermore, this framework has also been employed in the study of the gauged Thirring model \cite{ref23}, where
 the nonrenormalizability of the model and its dynamical mass generation were proved.
In all the above cases the analysis of the divergence nature of higher
order graphs has shown to be rather clear and stronger than the usual (naive)
power counting method.

It should be emphasized, however, that the study of normalizability for these models is possible because the causal
 approach follows an inductive method to construct each and every element of the S-matrix series. So, we can analyze the
  properties of a particular model for each one of these elements, in particular its singularity degree. 
  Subsequently, we can determine unambiguously how many constants are necessary to fix or normalize the given model.
In this work we shall use the causal method to study the normalizability of the GQED$_{4}$ and its consequence
onto its gauge invariance.
Also, the Ward-Takahashi-Fradkin identities are perturbatively determined.

Hence, the main aim of the this paper is to illustrate how the causal method can be employed in order
to study systematically the normalizability of a higher derivative field theory, in particular by discussing the subtle
points involving other methods, so that this distributional analysis, in face of standard naive methods,
could be generally applied to more phenomenologically field theories, with special interest in gravitational theories.
The work is organized as follows. In Sec.~\ref{sec:2}, we briefly review the
Bopp-Podolsky and Dirac fields, we also give some essentials
properties of these fields necessary to develop the inductive method.
In Sec.~\ref{sec:3}, we summarize the main points of the Epstein-Glaser method in order to establish
 the inductive method.
In Sec.~\ref{sec:4}, we demonstrate the normalizability of the GQED$_{4}$ by determining the singular
 order of each term of the perturbative S-matrix series. In Sec.~\ref{sec:5}, we use the previous 
 obtained results to show, from the causal point of view, the gauge invariance of the GQED$_{4}$ 
 and also that the Ward-Takahashi-Fradkin identities are consistently satisfied order-by-order.
Finally, our conclusions and remarks are given in Sec.~\ref{sec:6}.


\section{Podolsky's electrodynamics}

\label{sec:2}

The lepton-photon interaction can be described as well by the generalized quantum
electrodynamics (GQED$_{4}$), which is endowed with a local $U\left(
1\right) $ gauge invariance. The GQED$_{4}$ is described by the
Lagrangian%
\begin{equation}
\mathcal{L}_{GQED}=\mathcal{L}_{D}+\mathcal{L}_{P}+\mathcal{L}_{int}.
\end{equation}%
The quantities $\mathcal{L}_{D}$ and $\mathcal{L}_{P}$ are, respectively, the
free Lagrangians that describe the Dirac and Bopp-Podolsky
electromagnetic free fields and define their propagators, at last, $\mathcal{L}
_{int}$ is the interaction part.

Moreover, we have that the dynamics of the free Dirac fields 
$\left(\psi ,\bar{\psi}=\psi ^{\dag }\gamma _{0}\right)$ is governed by $\mathcal{L}_{D}=\bar{%
\psi}\left( i\gamma .\partial -m\right) \psi $. Hence, these fields satisfy
the free Dirac equations%
\begin{equation}
\left( i\gamma .\partial -m\right) \psi =0,\text{\quad }\bar{\psi}%
\left(i\gamma .\overleftarrow{\partial }+m\right) =0,  \label{DirEq}
\end{equation}%
where $m$ is the physical mass of the fermions. Recall that all the parameters are (asymptotically) 
physical within the causal approach.
Using the analytic representation for propagators \cite{ref14}, we can
find the positive (PF) and negative (NF) frequency fermionic propagators%
\begin{equation}
\hat{S}^{\left( \pm \right) }\left( p\right) =\left( \gamma .p+m\right) \hat{%
D}_{m}^{\left( \pm \right) }\left( p\right),
\end{equation}%
where $\hat{D}_{m}^{\left( \pm \right) }\left( p\right) =\pm \frac{i}{2\pi }
\theta \left( \pm p_{0}\right) \delta \left( p^{2}-m^{2}\right) $ are the PF
and NF massive scalar propagators. These propagators are related to the
contraction of the operator fermionic fields as follows%
\begin{equation}
\overbrace{\psi _{e}\left( x\right) \bar{\psi}_{f}\left( y\right) }=\frac{1}{%
i}S_{ef}^{\left( +\right) }\left( x-y\right), \quad \overbrace{\bar{\psi}%
_{e}\left( x\right) \psi _{f}\left( y\right) }=\frac{1}{i}%
S_{fe}^{\left(-\right) }\left( y-x\right) .
\end{equation}%
On the other hand, for the Bopp-Podolsky electromagnetic field we consider
that it is described by the following gauge-fixed Lagrangian \cite{ref14}%
\begin{equation}
\mathcal{L}_{P}=-\dfrac{1}{4}F_{\mu \nu }F^{\mu \nu }+\dfrac{a^{2}}{2}%
\partial _{\mu }F^{\mu \sigma }\partial ^{\nu }F_{\nu \sigma }-\dfrac{1}{%
2\xi }\left( \partial .A\right) \left( 1+a^{2}\square \right) \left(
\partial .A\right) ,
\end{equation}%
where $F_{\mu \nu }=\partial _{\mu }A_{\nu }-\partial _{\nu }A_{\mu }$ is
the field strength, $a$ is the Bopp-Podolsky parameter, and $\xi $ is the
constant gauge-fixing parameter. The gauge-fixing term was introduced via
the Lagrange multiplier method, where we have chosen to work with the \emph{%
non-mixing gauge condition} \cite{ref24}%
\begin{equation}
\left( 1+a^{2}\square \right) ^{1/2}\partial ^{\mu }A_{\mu }=0,
\label{Non-mix-cond}
\end{equation}%
which is a pseudodifferential equation \cite{ref25}. It is important to
emphasize that for the non-mixing gauge condition we can obtain, for the
choice $\xi =1$, the following field equation: 
\begin{equation}
\left( 1+a^{2}\square \right) \square A_{\mu }=0.
\end{equation}%
Hence, from this expression, we can conclude that the Bopp-Podolsky field has
two non-mixing sectors
\begin{align}
\square A_{\mu } ^M=0, \quad \left( \square +m_{a}^{2}\right) A_{\mu }^P =0,
\end{align}%
a massless and massive propagating modes (see \eqref{prop}), i.e. Maxwell and Proca sectors, respectively.
The mass of the photon in the Proca sector is given by: $m_{a}=a^{-1}$.
Further on, again using the analytic representation \cite{ref14}, we can find
the PF and NF electromagnetic propagators%
\begin{align}
\hat{D}_{\mu \nu }^{\left( \pm \right) }\left( k\right) &=g_{\mu \nu }\left( 
\hat{D}_{0}^{\left( \pm \right) }\left( k\right) -\hat{D}_{m_{a}}^{\left(
\pm \right) }\left( k\right) \right) -\left( 1-\xi \right) k_{\mu }k_{\nu }%
\hat{D}_{0}^{\prime \left( \pm \right) }\left( k\right) \nonumber \\
&\quad +\left( 1-\xi \right) \frac{k_{\mu }k_{\nu }}{m_{a}^{2}}\left( \hat{D}_{m_{a}}^{\left( \pm
\right) }\left( k\right) -\hat{D}_{0}^{\left( \pm \right) }\left( k\right)
\right) ,\label{prop}
\end{align}
where $\hat{D}_{0}^{\left( \pm \right) }$ and $\hat{D}_{m_{a}}^{\left( \pm
\right) }$ are the PF and NF scalar propagators for the massless and massive
modes, respectively; in particular, $\hat{D}_{0}^{\prime \left( \pm \right)
}\left(k\right) =\mp \frac{i}{2\pi }\theta \left( \pm k_{0}\right) \delta
\left( k^{2}\right) $ are the PF and NF dipolar massless scalar propagators.
The PF electromagnetic propagator is related to the following contraction of
the electromagnetic fields 
\begin{equation}
\overbrace{A_{\mu }\left( x\right) A_{\nu }\left( y\right) }\equiv \left[%
A_{\mu }^{\left( -\right) }\left( x\right) ,A_{\nu }^{\left( +\right)
}\left( y\right) \right] =iD_{\mu \nu }^{\left( +\right) }\left( x-y\right) .
\end{equation}
Finally, according to the minimal coupling principle, we have that the
interaction Lagrangian, $\mathcal{L}_{int}$, is given by%
\begin{equation}
\mathcal{L}_{int}=e:\bar{\psi}\left( x\right) \gamma ^{\mu }\psi \left(
x\right) :A_{\mu }\left( x\right) ,  \label{LINT}
\end{equation}%
where $:$ $:$ indicates the normal ordering and $e$ is the constant coupling
(normalized, in the causal approach). Also, we can identify $j^{\mu }\left(
x\right) =:\bar{\psi}\left( x\right) \gamma ^{\mu }\psi \left( x\right) :$
as the electromagnetic current.


\section{The S-matrix's inductive causal program}

\label{sec:3}

In this section, the construction of a perturbative quantum field theory is
reviewed from the point of view of the Epstein-Glaser causal theory \cite%
{ref16}. In this approach the S-matrix is constructed with no reference
to the Hamiltonian formalism, rather it consider an axiomatic formulation.
The causal approach postulates the S-matrix in the following formal
perturbative series 
\begin{equation}
S\left[ g\right] =1+\sum\limits_{n=1}^{\infty }\frac{1}{n!}\int
dx_{1}dx_{2}...dx_{n}T_{n}\left( x_{1},x_{2},...,x_{n}\right) g\left(
x_{1}\right) g\left( x_{2}\right) ...g\left( x_{n}\right) ,  \label{S}
\end{equation}%
where we can identify the quantity $T_{n}$ as an operator-valued
distribution and $g^{\otimes n}$ its test function. In order to guarantee the
existence of (momentum) Fourier transformed expressions it is considered that the
 test function $g$ belongs to the Schwartz space $\mathcal{J}\left( M^{4}\right) $. \footnote{%
The test function $g$ plays the part of switching on (off) the interaction,
thus $g\left( x\right) \in \left[ 0,1\right] $. Hence, when the limit $%
g\rightarrow 1$ is taken adiabatically, we have a free system.} In a similar
way, we have that the inverse S-Matrix has the form \cite{ref29} 
\begin{equation}
S^{-1}\left[ g\right] =1+\sum\limits_{n=1}^{\infty }\frac{1}{n!}\int
dx_{1}dx_{2}...dx_{n}\tilde{T}_{n}\left( x_{1},x_{2},...,x_{n}\right)
g\left( x_{1}\right) g\left( x_{2}\right) ...g\left( x_{n}\right) ,
\end{equation}%
where the distributions $\tilde{T}_{n}$ can be obtained by a formal
inversion of $\left( \ref{S}\right) $. Then, we find the relation \cite{ref21}%
\begin{equation}
\tilde{T}_{n}\left( X_{n}\right) =\sum_{r=1}^{n}\left( -1\right)
^{r}\sum_{P_{r}}\left[ T_{n_{1}}\left( X_{1}\right) ...T_{n_{r}}\left(
X_{r}\right) \right] ,  \label{Tinv}
\end{equation}%
where the sum runs over all partitions $P_{r}$ of $\left\{
x_{1},...,x_{n}\right\} $ into non-empty $r$ disjoint sets: $%
X_{n}=\bigcup\limits_{j=1}^{r}X_{j}$, with $\ X_{j}\neq \emptyset $ and $\left\vert
X_{j}\right\vert =n_{j}$.

In this axiomatic approach, the construction of the building blocks $T_{n}$
is given via the inductive method. This method is determined when we consider,
as postulates, the general physical principles of \emph{causality} \cite%
{ref27}, \emph{relativistic invariance} \cite{ref28}, and the \emph{%
asymptotic conditions} in the sense of Heisenberg's program \cite{ref17}. Since in this
approach the S-matrix is a functional of the test function $g$ \cite%
{ref29}, these postulates can easily be introduced as follows:

\begin{itemize}

\item [(i)]\emph{Causality}, this principle is understood as the possibility of
localizing and ordering events in the space-time. Thus, it can be formulated
by the causal ordering relation%
\begin{equation}
S\left[ g_{1}+g_{2}\right] =S\left[ g_{2}\right] S\left[ g_{1}\right] ,\quad 
\text{if} ~~\text{Supp}~\left( g_{1}\right) <\text{Supp}~\left( g_{2}\right)
,
\end{equation}%
when we substitute this into the perturbative S-matrix series %
\eqref{S}, we arrive into the causal relation for the $T_{n}$ distributions:%
\begin{equation}
T_{n}\left( x_{1},...,x_{m},x_{m+1},...,x_{n}\right) =T_{m}\left(
x_{1},...,x_{m}\right) T_{n-m}\left( x_{m+1},...,x_{n}\right) ,
\end{equation}%
if the inequality is satisfied $\left\{ x_{1},...,x_{m}\right\} >\left\{ x_{m+1},...,x_{n}\right\} $, 
\footnote{%
Which means that $x_{j}^{0}>x_{i}^{0}$, for $j=1,\ldots ,m$ and $%
i=m+1,\ldots ,n$.} then we can say that $T_{n}$ is a \emph{causal ordered
product} distribution. Since the sign $>$ is understood in \textit{stricto
sensu}, the distribution $T_{n}$ cannot be expressed in terms of the well
known Feynman time-ordering product: $T_{n}\left( x_{1},...,x_{n}\right)
\neq \mathcal{T}\left[T_{1}\left( x_{1}\right) \cdots T_{1}\left(
x_{n}\right) \right] $, which is endowed with UV divergences. \footnote{%
In the time-ordered products a Heaviside $\theta $-function is present, 
one can show that the product of this function with
singular distributions is in fact a divergent quantity \cite{ref26}.}

\item [(ii)]\emph{Relativistic invariance}, in general, $\mathcal{U}$ is a symmetry
if for two observers $\mathcal{O}$ and $\mathcal{O}^{\prime }$, which look to 
the same system, the measured transition probabilities are equal. If we consider a
 unique asymptotically free particle space $\mathcal{F}$, the symmetry $\mathcal{U}
$ can be represented by a single operator $U:\mathcal{F}\rightarrow \mathcal{%
F} ^{\prime }$. Then, for the S-matrix: $S$ and $S^{\prime }$ observed by $\mathcal{O}$ 
and $\mathcal{O}^{\prime }$ are, respectively, related as follows%
\begin{equation}
S^{\prime }=U~S~U^{-1}.
\end{equation}%
The Epstein-Glaser method considers that, for the deduction of each element
of the perturbative series \eqref{S}, it is sufficient to take as (symmetry) axioms the
translational and Lorentz invariance:%
\begin{align}
U\left( 1,a\right) ~S\left[ g\right] ~U^{-1}\left( 1,a\right) =&S\left[ g_{a}%
\right] , \quad g_{a}\left( x\right) =g\left( x-a\right) , \\
U\left( \Lambda ,0\right)~ S\left[ g\right] ~U^{-1}\left( \Lambda ,0\right)
=&S\left[ g_{\Lambda }\right] , \quad g_{\Lambda }\left( x\right) =g\left(
\Lambda ^{-1}x\right) ,
\end{align}%
where $U\left( \Lambda ,a\right) $ is the continuous unitary representation
of the orthochronous Poincar\'{e} group. Furthermore, when we replace these
relations into the series \eqref{S}, we find that the operator-valued
distributions $T_{n}$ can be written as
\begin{equation}
T_{n}\left( x_{1},x_{2},...,x_{n}\right)
=T_{n}\left(x_{1}-x_{n},x_{2}-x_{n},...,x_{n-1}-x_{n}\right) ,
\end{equation}
whereas the Lorentz invariance implies that
\begin{equation}
U~\left( \Lambda ,0\right) T_{n}\left( x_{1},x_{2},...,x_{n}\right)~
U^{-1}\left( \Lambda ,0\right) =T_{n}\left( \Lambda x_{1},\Lambda
x_{2},...,\Lambda x_{n}\right) .
\end{equation}

\item [(iii)]\emph{Asymptotic conditions and interaction}, in this formalism only the
free asymptotic fields acting on the Fock space $\mathcal{F}$ are used in
order to construct $S\left[ g\right] $. Thus, for the GQED$_{4}$,
we shall consider the set of electromagnetic and spinor free fields: $
\left(A_{\mu }, \psi ,\bar{\psi}\right) $. This axiom also says that the one-point
 distribution $T_{1}\left( x\right) $ is proportional to the interaction
Lagrangian. Thus, from Eq.\eqref{LINT}, we have that%
\begin{equation}
T_{1}\left( x\right) =ie:\bar{\psi}\left( x\right) \gamma ^{\mu }\psi \left(
x\right) :A_{\mu }\left( x\right) ,
\end{equation}%
$e$ is the normalized constant coupling.
\end{itemize}

The inductive method starts with the initial data $T_{1}$ and also with
 $\tilde{T}_{1}$ (due to Eq.\eqref{Tinv}). Then, from these initial data, we can find
the $2$-point distribution $T_{2}$. In general, the inductive method
proposes to find the $n$-order term $T_{n}$ from the set $\left\{
T_{1},\ldots ,T_{n-1},\tilde{T}_{1},\ldots ,\tilde{T}_{n-1}\right\} $. For
this purpose, the Epstein-Glaser approach introduce a well-defined
distributional product, such as the intermediate $n$-point distributions 
\begin{align}
A_{n}^{\prime }\left( x_{1},...,x_{n}\right) \equiv &\sum_{P_{2}}\tilde{T}%
_{n_{1}}\left( X\right) T_{n-n_{1}}\left( Y,x_{n}\right) ,  \label{Alin} \\
R_{n}^{\prime }\left( x_{1},...,x_{n}\right) \equiv
&\sum_{P_{2}}T_{n-n_{1}}\left( Y,x_{n}\right) \tilde{T}_{n_{1}}\left(
X\right) ,  \label{Rlin}
\end{align}%
where $P_{2}$ are all partitions of $\left\{ x_{1},...,x_{n-1}\right\} $
into the disjoint sets $X,$ $Y$ such that $\left\vert X\right\vert
=n_{1}\geq 1$ and $\left\vert Y\right\vert \leq n-2$. Moreover, other
important distributions are obtained when the sums in Eqs.\eqref{Alin} and %
\eqref{Rlin} are now extended over all partitions $P_{2}^{0}$, including the
 empty set. These are the advanced and retarded distributions%
\begin{align}
A_{n}\left( x_{1},...,x_{n}\right) \equiv &\sum_{P_{2}^{0}}\tilde{T}%
_{n_{1}}\left( X\right) T_{n-n_{1}}\left( Y,x_{n}\right)  \label{Adef} \\
=&A_{n}^{\prime }\left( x_{1},...,x_{n}\right)
+T_{n}\left(x_{1},...,x_{n}\right) ,  \nonumber \\
R_{n}\left( x_{1},...,x_{n}\right) \equiv &\sum_{P_{2}^{0}}T_{n-n_{1}}\left(
Y,x_{n}\right) \tilde{T}_{n_{1}}\left( X\right)  \label{Rdef} \\
=&R_{n}^{\prime }\left( x_{1},...,x_{n}\right) +T_{n}\left(
x_{1},...,x_{n}\right) .  \nonumber
\end{align}%
By causal properties, one may easily conclude that $R_{n}$ and $A_{n}$ have
retarded and advanced support, respectively,%
\begin{equation}
\text{Supp}~R_{n}\left( x_{1},...,x_{n}\right) \subseteq \Gamma
_{n-1}^{+}\left( x_{n}\right) , \quad \text{Supp}~A_{n}\left(
x_{1},...,x_{n}\right) \subseteq \Gamma _{n-1}^{-}\left( x_{n}\right) ,
\end{equation}%
where $\Gamma _{n-1}^{\pm }\left( x_{n}\right) =\left\{
\left(x_{1},...,x_{n}\right) /~~x_{j}\in \overline{V}^{\pm }\left(
x_{n}\right) ,\quad \forall ~j=1,...,n-1\right\} $, and $\overline{V}^{\pm }\left(
x_{n}\right) $ is the closed forward (backward) cone. These two
distributions are not determined by the induction assumption, rather they are
obtained by the \emph{splitting} process of the so-called \emph{causal
distribution} defined as 
\begin{equation}
D_{n}\left( x_{1},...,x_{n}\right) \equiv R_{n}^{\prime }\left(
x_{1},...,x_{n}\right) -A_{n}^{\prime }\left( x_{1},...,x_{n}\right)
=R_{n}\left( x_{1},...,x_{n}\right) -A_{n}\left( x_{1},...,x_{n}\right) .
\label{Ddef}
\end{equation}%
For the case of GQED$_{4}$ we can write $D_{n}$ as follows%
\begin{equation}
D_{n}\left( x_{1},...,x_{n}\right)
=\sum\limits_{k}d_{n}^{k}\left(x_{1},...,x_{n}\right) :\prod\limits_{j}\bar{%
\psi}\left( x_{j}\right) \prod\limits_{l}\psi \left( x_{l}\right)
\prod\limits_{m}A\left(x_{m}\right) :,
\end{equation}%
where $d_{n}^{k}\left( x_{1},...,x_{n}\right) $ is the numerical part of the
distribution. Besides, from the translational invariance, we see that $d_{n}^{k}$
depends only on relative coordinates:%
\begin{equation}
d\left( x\right) \equiv d_{n}^{k}\left(
x_{1}-x_{n},...,x_{n-1}-x_{n}\right)~~ \in \mathcal{J}^{\prime }\left( 
\mathbb{R}^{m}\right) ,\quad m=4\left( n-1\right) .
\end{equation}%
As aforementioned an important step in the analysis is the splitting of the
numerical causal distribution $d$ at origin: $\left\{ x_{n}\right\}
=\Gamma _{n-1}^{+}\left( x_{n}\right) \cap \Gamma _{n-1}^{-}\left(
x_{n}\right) $, into the advanced and retarded parts. These distributions are
 denoted as $a$ and $r$, respectively. When we analyse the convergence of the
sequence $\left\{ \left\langle d,\phi _{\alpha }\right\rangle \right\} $,
where $\phi _{\alpha }$ has a decreasing support when $\alpha \rightarrow
0^{+} $ and belongs to the Schwartz space $\mathcal{J}$, we find some
natural distributional definitions. For instance, we name $d$ as a
distribution of singular order $\omega $ if its Fourier transform $\hat{d}%
\left( p\right) $ has a quasi-asymptotic $\hat{d}_{0}\left( p\right) \neq 0$
at $p=\infty $ with regard to a positive continuous function $\rho \left(
\alpha \right) $, $\alpha >0$, i.e. if the limit%
\begin{equation}
\lim_{\alpha \rightarrow 0^{+}}\rho \left( \alpha \right) \left\langle \hat{d%
}\left( \frac{p}{\alpha }\right) ,\phi \left( p\right) \right\rangle
=\left\langle \hat{d}_{0}\left( p\right) ,\phi \left( p\right) \right\rangle
\neq 0,  \label{LIM1}
\end{equation}%
exists in $\mathcal{J}^{\prime }\left( \mathbb{R}^{m}\right) $, with the 
\textit{power-counting} function $\rho \left( \alpha \right) $ satisfying 
\begin{equation}
\lim_{\alpha \rightarrow 0}\frac{\rho \left( a\alpha \right) }{\rho
\left(\alpha \right) }=a^{\omega },\quad \forall ~~a>0,
\end{equation}%
or, equivalently, 
\begin{equation}
\rho \left( \alpha \right) \rightarrow \alpha ^{\omega }L\left( \alpha
\right),~\text{when}~\alpha \rightarrow 0^{+},
\end{equation}%
where $L\left( \alpha \right) $ is a quasi-constant function at $\alpha =0$.
Of course, there is an equivalent definition in the coordinate space, but,
since the splitting process is more easily accomplished in the momentum
space, this one suffices for our purposes. From this definition we have two
distinct cases depending on the value of $\omega $ \cite{ref21}, these
are:

\textit{(i) }\emph{Regular} distributions - for $\omega <0$, in this case
the solution of the splitting problem is unique and the retarded
distribution is defined by multiplying $d$ by step functions, its form in
the momentum space is given as follows%
\begin{equation}
\hat{r}\left( p\right) =\frac{i}{2\pi }sgn\left( p^{0}\right) \int_{-\infty
}^{+\infty }dt\frac{\hat{d}\left( tp\right) }{\left(
1-t+sgn\left(p^{0}\right) i0^{+}\right) },
\end{equation}%
identified as a dispersion relation without subtractions.

\textit{(ii) }\emph{Singular} distributions - for $\omega \geq 0$, then the
solution cannot be obtained as in the \emph{regular} case and, after a
careful mathematical treatment, it may be shown that the retarded
distribution is given by the central splitting solution%
\begin{equation}
\hat{r}\left( p\right) =\frac{i}{2\pi }sgn\left( p^{0}\right) \int_{-\infty
}^{+\infty }dt\frac{\hat{d}\left( tp\right) }{t^{\omega
+1}\left(1-t+sgn\left( p^{0}\right) i0^{+}\right) },
\end{equation}%
identified as a dispersion relation with $\omega +1$ subtractions. But in
contrast with the regular case, this solution is not unique. If $\hat{r}$ is
a retarded part, then $\tilde{r}$ defined as 
\begin{equation}
\tilde{r}\left( p\right) =\hat{r}\left( p\right) +\sum_{a=0}^{\omega
}C_{a}p^{a},
\end{equation}%
is also a retarded part. The undetermined constants are not fixed by
causality, and any other condition already introduced here, rather
additional physical ``normalization'' conditions are necessary to fix them.

It is important to emphasize, however, that by the definition of the causal
 distribution \eqref{Ddef}, the singular order of the intermediate distributions, retarded and advanced parts, are in fact equal:%
\begin{equation}
\omega \left( D\right) =\omega \left( R^{\prime }\right) =\omega \left(
A^{\prime }\right) =\omega \left( R\right) =\omega \left( A\right) .
\label{EqSinOrd}
\end{equation}%

Moreover, within some particular contexts, it shows to be useful to introduce the concept
of graph in the causal approach. First, it should be clear that in this approach
these are not Feynman integrals, the graphs are actually a purely schematic tool,
so they do not play any part in the calculation itself.
A class of graph must be understood here as the set of all elements of the S-matrix
series \eqref{S} so that we have the same configuration.
This means that, in our case, a particular class is such that the graphs have the
same number of non-contracted spinor fields and/or the same number of non-contracted electromagnetic
fields. Moreover, a given graph is seen as an individual contribution.
Then, the $n $-order distribution $T_{n}$ can be written as the sum of $n$-order graphs of different classes%
\begin{equation}
T_{n}=\sum\limits_{g}T_{n}^{g}\left( x_{1},x_{2},...,x_{n}\right) ,
\end{equation}%
in this expression $T_{n}^{g}$ contains graphs of the same class, and, for
GQED$_{4}$, each one of them can be written as follows%
\begin{equation}
T_{n}^{g}\left( x\right) =\sum \colon \prod\limits_{j=1}^{f_{g}}\bar{\psi}%
\left( x_{k_{j}}\right) t_{g}\left( x_{1},x_{2},...,x_{n}\right)
\prod\limits_{j=1}^{f_{g}}\psi \left( x_{n_{j}}\right) \colon
:\prod\limits_{j=1}^{l_{g}}A_{\mu _{m_{j}}}\left( x_{m_{j}}\right) :,
\end{equation}%
where $l_{g}$ indicates the number of non-contracted electromagnetic fields, 
$2f_{g}$ is the number of non-contracted spinors fields, $t_{g}\left(
x_{1},x_{2},...,x_{n}\right) $ is the contracted or numerical part and the
sub-index $g$ indicates a given fixed configuration. In particular, we define some important graphs:
\begin{enumerate}
\item The $ n $-point lepton-lepton scattering graphs,
\begin{equation}
T_{n}^{LL}\left( x\right) =\sum \colon \prod\limits_{j=1}^{2}\bar{\psi}%
\left( x_{k_{j}}\right) t_{g}\left( x_{1},x_{2},...,x_{n}\right)
\prod\limits_{j=1}^{2}\psi \left( x_{n_{j}}\right) \colon ;
\end{equation}
\item The $ n $-point lepton-photon scattering graphs,
\begin{equation}
T_{n}^{LP}\left( x\right) =\sum \colon\bar{\psi}%
\left( x_{k_{1}}\right) t_{g}\left( x_{1},x_{2},...,x_{n}\right)
\psi \left( x_{n_{1}}\right) \colon
\colon\prod\limits_{j=1}^{2}A_{\mu _{m_{j}}}\left( x_{m_{j}}\right) \colon;
\end{equation}
\item The $ n $-point (fermionic) self-energy graphs,
\begin{equation}
T_{n}^{SE}\left( x\right) =\sum \colon\bar{\psi}%
\left( x_{k_{1}}\right) t_{g}\left( x_{1},x_{2},...,x_{n}\right)
\psi \left( x_{n_{1}}\right) \colon ;
\end{equation}
\item The $ n $-point vacuum polarization graphs,
\begin{equation}
T_{n}^{VP}\left( x\right) =\sum \colon A_{\mu _{m_{1}}}\left( x_{m_{1}}\right)
 t_{g}\left( x_{1},x_{2},...,x_{n}\right) A_{\mu _{m_{2}}}\left( x_{m_{2}}\right) \colon.
\end{equation}
\end{enumerate}


\section{Normalizability}

\label{sec:4}

The definition of normalizability in the causal perturbation theory is closely
related to the singular order of each graph that contributes to the S-matrix
series \eqref{S}. In the previous section we have mentioned that if the
retarded part is a singular distribution then it is not uniquely defined.
This implies that for every graph of $n$-order $T_{n}^{g}\left( x\right) $,
that has a singular order $\omega \geq 0$, a polynomial of degree $\omega $
shall remain undetermined in momentum space. Hence, in each graph of $n$%
-order, with a finite number of free parameters, there are three
possibilities when considering the inductive procedure:

\begin{itemize}

\item[(i)] The number of free parameters increases with $n$ without bound;
then the model is called non-normalizable.

\item[(ii)] The total number of free parameters appearing in each order is
finite; then the model is normalizable.

\item[(iii)] There is only a finite number of low-order graph with $\omega
\geq 0$; then the model is called super-normalizable.
\end{itemize}

In order to determine the normalizability of the GQED$_4$, i.e. to determine
the singular order of an arbitrary graph, we must first calculate the
singular order of the contraction of two graphs via the inductive method.


\subsection{Singular order of two contracted graphs}

Following the inductive construction, we define the numerical part of a $r$%
-point graph $G_{1}$ and a $s$-point graph $G_{2}$ by the distributions%
\begin{equation}
t_{1}\left( x_{1}-x_{r},\cdots ,x_{r-1}-x_{r}\right) ,\quad t_{2}\left(
y_{1}-y_{s},\cdots ,y_{s-1}-y_{s}\right) ,
\end{equation}%
respectively. Besides, the graph that contributes to the S-matrix must be
obtained after the splitting procedure, and then we can finally determine its singular
order. But, as we have already mentioned in Sect.~\ref{sec:3} (see Eq.\eqref{EqSinOrd}),
 a graph has the same singular order of a contracted
graph (intermediate distribution formed by contractions). Therefore, for
our purposes, it is enough to calculate the singular order of a contracted graph.

For instance, if we consider that the graphs $G_{1}$ and $G_{2}$ are
contracted by $\kappa$-contractions of a same class of field; then, by
taking translational invariance into account, the numerical part of the
contracted graph takes the form:%
\begin{equation}
t_{1}\left( x_{1}-x_{r},\cdots ,x_{r-1}-x_{r}\right)
\prod\limits_{j=1}^{\kappa }D_{a_{r_{j}}b_{s_{j}}}^{\left( +\right) }\left(
x_{r_{j}}-y_{s_{j}}\right) t_{2}\left( y_{1}-y_{s},\cdots
,y_{s-1}-y_{s}\right) ,  \label{ContrGr1}
\end{equation}%
where $D_{a_{r_{j}}b_{s_{j}}}^{\left( +\right) }$ indicates the numerical
part of the contraction between the points $x_{r_{j}}$ and $y_{s_{j}}$.
Also, $a_{r_{j}}$ and $b_{s_{j}}$ are the index of the associated fields into
these points, respectively.

Moreover, we can write the numerical part of the contracted graph %
\eqref{ContrGr1} using the relative variables%
\begin{align}
\xi _{i} =&x_{i}-x_{r}, \quad i=1,\ldots ,r-1, \\
\eta _{k} =&y_{k}-y_{s}, \quad k=1,\ldots ,s-1, \\
\eta =&x_{r}-y_{s},
\end{align}%
so that we obtain 
\begin{equation}
t\left( \xi _{1},\cdots ,\xi _{r-1},\eta _{1},\cdots ,\eta _{s-1},\eta
\right) =t_{1}\left( \xi _{1},\cdots ,\xi _{r-1}\right)
\prod\limits_{j=1}^{\kappa }D_{a_{r_{j}}b_{s_{j}}}^{\left( +\right) }\left(
\xi _{r_{j}}-\eta _{s_{j}}+\eta \right) t_{2}\left( \eta _{1},\cdots ,\eta
_{s-1}\right) .
\end{equation}%
In order to compute the singular order of the contracted graph, we evaluate
the Fourier transform of its numerical part, \footnote{%
A detailed calculation can be found in the Appendix \ref{Sec:B}.} which is
given by the convolution%
\begin{align}
\hat{t}\left( p_{1},\cdots ,p_{r-1},q_{1},\cdots ,q_{s-1},q\right) &=\hat{t}
_{1}\left( \cdots ,p_{i},\cdots ,p_{r_{j}}-k_{j},\cdots \right) \hat{t}%
_{2}\left( \cdots ,q_{k},\cdots ,q_{s_{j}}+k_{j},\cdots \right)  \notag \\
&\quad \times \left( 2\pi \right) ^{-\frac{4\kappa }{2}}\int
\prod\limits_{j=1}^{\kappa }dk_{j}\delta \left( q-\sum_{j}k_{j}\right)
\prod\limits_{j=1}^{\kappa }\hat{D}_{a_{r_{j}}b_{s_{j}}}^{\left( +\right)
}\left( k_{j}\right) .
\end{align}
Now we can calculate the power counting function $\rho $ of the contracted
graph by using the definition of the limit \eqref{LIM1}. Moreover, we need to
compute the distribution $\hat{t}\left( p\right) $, which has the
dimension $m=4\left( r+s-1\right) $, evaluated in the test function $\check{%
\varphi}\left( \alpha p\right) $ ,
\begin{align}
\left\langle \hat{t}\left( p\right) ,\check{\varphi}\left( \alpha p\right)
\right\rangle &=\int d^{r-1}pd^{s-1}qdq\hat{t}_{1}\left( \cdots
,p_{i},\cdots ,p_{r_{j}}-k_{j},\cdots \right) \hat{t}_{2}\left( \cdots
,q_{k},\cdots ,q_{s_{j}}+k_{j},\cdots \right)  \notag \\
&\quad \times \left( 2\pi \right) ^{-\frac{4\kappa }{2}}\int
\prod\limits_{j=1}^{\kappa }dk_{j}\delta \left( q-\sum_{j}k_{j}\right)
\prod\limits_{j=1}^{\kappa }\hat{D}_{a_{r_{j}}b_{s_{j}}}^{\left( +\right)
}\left( k_{j}\right)  \notag \\
&\quad \times \check{\varphi}\left( \cdots ,\alpha p_{i},\cdots ,\alpha
p_{r_{j}},\cdots ,\alpha q_{k},\cdots ,\alpha q_{s_{j}},\cdots ,\alpha
q\right) .
\end{align}
It shows to be convenient for calculation purposes to consider the following
change of variables%
\begin{equation}
\alpha p_{i}\rightarrow p_{i}^{\prime },\quad \alpha p_{r_{j}}\rightarrow
p_{r_{j}}^{\prime } +k_{j},\quad \alpha q_{k}\rightarrow q_{k}^{\prime },
\quad \alpha q_{s_{j}}\rightarrow q_{s_{j}}^{\prime }- k_{j}, \quad \alpha
q\rightarrow q^{\prime },  \label{Varex}
\end{equation}%
thus, after considering carefully this change into the integrating
variables, we obtain that 
\begin{align}
\left\langle \hat{t}\left( p\right) ,\check{\varphi}\left( \alpha p\right)
\right\rangle &=\int \frac{d^{r-1}p^{\prime }}{\alpha ^{4\left( r-1\right) }}%
\frac{d^{\nu -1}q^{\prime }}{\alpha ^{4\left( s-1\right) }}\frac{dq^{\prime }%
}{\alpha ^{4}}\hat{t}_{1}\left( \cdots ,\frac{p_{i}^{\prime }}{\alpha }%
,\cdots ,\frac{p_{r_{j}}^{\prime }}{\alpha }\text{\ },\cdots \right) \hat{t}%
_{2}\left( \cdots ,\frac{q_{k}^{\prime }}{\alpha }\text{ },\cdots ,\frac{%
q_{s_{j}}^{\prime }}{\alpha },\cdots \right)  \notag \\
&\quad\times \left( 2\pi \right) ^{-\frac{4\kappa }{2}}\int \frac{1}{\alpha
^{4\kappa }}\prod\limits_{j=1}^{\kappa }dk_{j}\delta \left( \frac{q^{\prime
}-\sum_{j}k_{j}}{\alpha }\right) \prod\limits_{j=1}^{\kappa }\hat{D}%
_{a_{r_{j}}b_{s_{j}}}^{\left( +\right) }\left( \frac{k_{j}}{\alpha }\right) 
\notag \\
&\quad\times \check{\varphi}\left( \cdots ,p_{i}^{\prime },\cdots
,p_{r_{j}}^{\prime }+k_{j},\cdots ,\cdots ,q_{k}^{\prime },\cdots
,q_{s_{j}}^{\prime }-k_{j},\cdots ,q^{\prime }\right) .  \label{Contra3}
\end{align}
Taking into account that in $4$-dimensions: $\delta \left( \frac{k}{\alpha }%
\right) =\alpha ^{4}\delta \left( k\right) $, and also the definition:%
\begin{equation}
\lim_{\alpha \rightarrow 0^{+}}\hat{D}_{a_{r_{j}}b_{s_{j}}}^{\left( +\right)
}\left( \frac{k}{\alpha }\right) =\alpha ^{-\sigma }\hat{D}%
_{a_{r_{j}}b_{s_{j}}}^{0\left( +\right) }\left( k\right),
\end{equation}%
where $\hat{D}_{a_{r_{j}}b_{s_{j}}}^{0\left( +\right) }$ is a nonvanishing
distribution and $\sigma $ is the singular order of $\hat{D}%
_{a_{r_{j}}b_{s_{j}}}^{\left( +\right) }$.
Furthermore, introducing the power counting
 functions $\rho _{1}\left( \alpha \right)$ and $\rho _{2}\left( \alpha \right) $
  (associated to $t_{1}$ and $t_{2}$, respectively) into Eq.\eqref{Contra3}, we arrive at:%
\begin{align}
&\left[ \alpha ^{-4}\rho _{1}\left( \alpha \right) \rho _{2}\left( \alpha
\right) \right] \alpha ^{m}\left\langle \hat{t}\left( p\right) ,\check{%
\varphi}\left( \alpha p\right) \right\rangle =  \notag \\
&=\frac{1}{\alpha ^{\left( 4+\sigma \right) \kappa }}\int d^{r-1}p^{\prime
}d^{s-1}q^{\prime }dq^{\prime }  \notag \\
&\times \left[ \rho _{1}\left( \alpha \right) \hat{t}_{1}\left( \cdots , 
\frac{p_{i}^{\prime }}{\alpha },\cdots ,\frac{p_{r_{j}}^{\prime }}{\alpha }%
\text{\ },\cdots \right) \right] \left[ \rho _{2}\left( \alpha \right) \hat{t%
}_{2}\left( \cdots ,\frac{q_{k}^{\prime }}{\alpha },\cdots ,\frac{%
q_{s_{j}}^{\prime }}{\alpha },\cdots \right) \right]  \notag \\
&\times \left( 2\pi \right) ^{-\frac{4\kappa }{2}}\int
\prod\limits_{j=1}^{\kappa }dk_{j}\delta \left(
q^{\prime}-\sum_{j}k_{j}\right) \prod\limits_{j=1}^{\kappa }\hat{D}%
_{a_{r_{j}}b_{s_{j}}}^{\left( +\right) }\left( k\right)  \notag \\
&\times \check{\varphi}\left( \cdots ,p_{i}^{\prime },\cdots
,p_{r_{j}}^{\prime }+k_{j},\cdots ,\cdots ,q_{k}^{\prime },\cdots
,q_{s_{j}}^{\prime }-k_{j},\cdots ,q^{\prime }\right) .  \label{Contra4}
\end{align}
Now we consider that the limits associated with the power counting functions of
the graphs $G_{1}$ and $G_{2}$ exist and that they are given by: $%
\lim\limits_{\alpha \rightarrow 0^{+}}\rho \left( \alpha \right) \hat{t}%
_{i}\left( \frac{p}{\alpha }\right) =\hat{t}_{i,0}\left( p\right) $. Then,
we can take the limit: $\alpha \rightarrow 0^{+}$ into the expression %
\eqref{Contra4},%
\begin{align}
&\lim_{\alpha \rightarrow 0^{+}}\left[ \alpha ^{-4}\rho _{1}\left( \alpha
\right) \rho _{2}\left( \alpha \right) \right] \alpha ^{m}\left\langle \hat{t%
}\left( p\right) ,\check{\varphi}\left( \alpha p\right) \right\rangle = 
\notag \\
&=\alpha ^{-\left( 4+\sigma \right) \kappa }\int d^{r-1}p^{\prime
}d^{s-1}q^{\prime }dq^{\prime }\hat{t}_{1,0}\left( \cdots ,p_{i}^{\prime
},\cdots ,p_{r_{j}}^{\prime },\cdots \right) \hat{t}_{2,0}\left(\cdots
,q_{k}^{\prime },\cdots ,q_{s_{j}}^{\prime },\cdots \right)  \notag \\
&\times \left( 2\pi \right) ^{-\frac{4\kappa }{2}}\int
\prod\limits_{j=1}^{\kappa }dk_{j}\delta \left( q^{\prime
}-\sum_{j}k_{j}\right) \prod\limits_{j=1}^{\kappa }\hat{D}
_{a_{r_{j}}b_{s_{j}}}^{0\left( +\right) }\left( k\right)  \notag \\
&\times \check{\varphi}\left( \cdots ,p_{i}^{\prime },\cdots
,p_{r_{j}}^{\prime }+k_{j},\cdots ,\cdots ,q_{k}^{\prime },\cdots
,q_{s_{j}}^{\prime }-k_{j},\cdots ,q^{\prime }\right) .
\end{align}
Besides, let us rewrite the above expression in terms of the original
variables \eqref{Varex}, we have that the limit
\begin{align}
&\lim_{\alpha \rightarrow 0}\left[ \alpha ^{\left( 4+\sigma \right) \kappa
-4}\rho _{1}\left( \alpha \right) \rho _{2}\left( \alpha \right) \right] 
\hat{t}\left( \cdots ,\frac{p_{i}}{\alpha },\cdots ,\frac{q_{k}}{\alpha }%
\text{ },\cdots ,\frac{q}{\alpha }\right) =  \notag \\
&=\left( 2\pi \right) ^{-\frac{4\kappa }{2}}\int \prod\limits_{j=1}^{\kappa
}dk_{j}\hat{t}_{1,0}\left( \cdots ,p_{i},\cdots ,p_{r_{j}}-k_{j},\cdots
\right) \hat{t}_{2,0}\left( \cdots ,q_{k}\text{ },\cdots
,q_{s_{j}}+k_{j},\cdots \right)  \notag \\
&\times \delta \left( q^{\prime }-\sum_{j}k_{j}\right)
\prod\limits_{j=1}^{\kappa }\hat{D}_{a_{r_{j}}b_{s_{j}}}^{0\left( +\right)
}\left( k\right) ,  \label{eq1}
\end{align}
exist in this sense. Furthermore, by comparing \eqref{eq1} with the definition \eqref{LIM1} 
\begin{equation}
\lim\limits_{\alpha \rightarrow 0^{+}}\rho \left( \alpha \right) \hat{t}%
\left( \frac{p}{\alpha }\right) =\hat{t}_{0}\left( p\right) ,
\end{equation}
we can see that the power counting function $\rho \left( \alpha \right) $
satisfies%
\begin{equation}
\rho \left( \alpha \right) \sim \alpha ^{\left( 4+\sigma \right) \kappa
-4}\rho _{1}\left( \alpha \right) \rho _{2}\left( \alpha \right) \sim \alpha
^{\left( 4+\sigma \right) \kappa -4+\omega \left( t_{1}\right) +\omega
\left( t_{2}\right) }.  \label{singularor}
\end{equation}%
Finally, we can affirm that if the singular order of the graphs $G_{1}$ and $%
G_{2}$ are $\omega \left( G_{1}\right) $ and $\omega \left( G_{1}\right) $,
respectively, then the singular order of the contracted graph $\omega \left(
G\right) $ can be identified from \eqref{singularor}  as given by
\begin{equation}
\omega \left( G\right) =\omega \left( G_{1}\right) +\omega
\left(G_{2}\right) +\left( 4+\sigma \right) \kappa -4.
\end{equation}%
recalling that $\kappa $ is the number of contracted fields, and $\sigma $
is the singular order of a single contraction. A simple generalization can
be obtained when this process is applied to the contraction of different kind
of quantum fields. Thus, in general, we find
\begin{equation}
\omega \left( G\right) =\omega \left( G_{1}\right) +\omega \left(
G_{2}\right) +\sum_{j}\left( 4+\sigma _{j}\right) \kappa _{j}-4,
\end{equation}%
where $\kappa _{j}$ is the number of contractions of the type $j$ with
singular order $\sigma _{j}$ associated to a particular class of quantum
field.

For GQED$_{4}$ we have two types of contractions:

\emph{(i)} The fermionic whose contractions $\hat{D}_{a_{r_{j}}b_{s_{j}}}^{1/2\left(+\right) }=S^{\left( \pm\right) }$
have singular order $\omega \left( S^{\left( \pm \right) }\right) =-1$;

\emph{(ii)} The electromagnetic whose contraction $\hat{D}%
_{a_{r_{j}}b_{s_{j}}}^{\left( +\right) }=\hat{D}_{\mu _{r_{j}}\nu
_{s_{j}}}^{\left( +\right) }$ has singular order $\omega \left( \hat{D}_{\mu
_{r_{j}}\nu _{s_{j}}}^{\left( +\right) }\right) =-4$.

Thus, if the GQED$_{4}$ graphs $G_{1}$ and $G_{2}$ (with singular order $%
\omega \left( G_{1}\right) $ and $\omega \left( G_{1}\right) $,
respectively) are contracted by $F$ spinorial and $L$
electromagnetic contractions, we have that the contracted graph $G$ has
singular order: 
\begin{equation}
\omega \left( G\right) =\omega \left( G_{1}\right) +\omega \left(
G_{2}\right) +3F-4.  \label{Contrafor}
\end{equation}%
It should be remarked that the difference between the singular order of the
electromagnetic propagator of QED$_{4}$ and GQED$_{4}$ is rather relevant, because
in GQED$_{4}$ the singular order $\omega \left( \hat{D}_{\mu \nu }\right) $
cancels exactly the value of the dimension frame considered \cite{ref14}.
\footnote{A similar analysis had been presented to show the renormalizability for higher-derivative
quantum gravity \cite{ref4}.}
This is exactly the reason why the singular order of the contracted graph,
in GQED$_{4}$, does not depend on the number of electromagnetic contractions.


\subsection{Super-normalizability proof}

Now we want to compute the singular order of any graph of the GQED$_4$.
The singular order of a graph is defined as the power of
its associated power counting function and, in general, a graph can be
obtained by the contraction of other graphs. Since the singular order of a
contraction is given by a linear combination (see Eq.\eqref{Contrafor}), we can
suppose that the singular order of a graph can be given by a linear
combination of the number of non-contracted spinor fields $f$ and the number
of non-contracted electromagnetic fields $l$, and that it also depends on the
perturbative order $n$, in the following form
\begin{equation}
\omega _{n}=an+bf+cl+d,  \label{HypSing}
\end{equation}%
where $a$, $b$, $c$, and $d$ are constants that we must determine. If this
argument is valid, then the graphs with $n_{1,2}$-point, $f_{1,2}$
non-contracted fermionic and $l_{1,2}$ non-contracted electromagnetic
fields have the singular order:%
\begin{align}
\omega _{n_{1}} =&an_{1}+bf_{1}+cl_{1}+d, \\
\omega _{n_{2}} =&an_{2}+bf_{2}+cl_{2}+d,
\end{align}
respectively. From the result \eqref{Contrafor} we can affirm that the
singular order of the contracted graph of $n=n_{1}+n_{2}$ order, with $F$
fermionic contractions and $L$ electromagnetic contractions, is given by%
\begin{equation}
\omega _{n}=an+b\left( f_{1}+f_{2}\right) +c\left( l_{1}+l_{2}\right)
+2d+3F-4,
\end{equation}%
also, by a hypothesis presented in \eqref{HypSing}, the singular order is equal to%
\begin{equation}
\omega _{n}=an+b\left( f_{1}+f_{2}-2F\right) +c\left( l_{1}+l_{2}-2L\right)
+d.
\end{equation}%
By comparing these two relations we have that $d=4$, $b=-\frac{3}{2}$ and $c=0$.
Then the singular order of the contracted graph is determined as
\begin{equation}
\omega _{n}=4+an-\frac{3}{2}f .
\end{equation}%
In order to fix the constant $a$ we need to consider a known case.
For this purpose, we may consider the results from Ref.~\cite{ref14}, where
we have found that the graph associated to the Bhabha scattering, i.e. $f=4$ and $n=2$,
 has singular order $-4$. Hence, we can conclude that $a=-1$. Finally, we find that
  the singular order of a $n$-point graph with $f$ and $l$
non-contracted spinor and electromagnetic fields, respectively, has the
singular order: \footnote{%
There are interesting differences of the present result in comparison with
the usual naive method \cite{ref32}, mainly because it does not depend on the
external electromagnetic lines. The usual approach is reviewed in the
appendix \ref{Sec:A}.} 
\begin{equation}
\omega _{n}=4-n-\dfrac{3}{2}f.  \label{SIngOrdFor}
\end{equation}%
A direct conclusion of this result is that for graphs with more than $5$%
-point the singular order is negative. Therefore, from our initial definition,
we can conclude that GQED$_{4}$ is a \emph{super-normalizable electrodynamics model}. 
Furthermore, we can determine the singular order of the lowest-order graphs:

\begin{enumerate}
\item The $4$-point graphs have singular order: $\omega _{4}=-\dfrac{3}{2}f$%
. Hence, the light-light scattering graph, the vacuum polarization graphs
and the $4$-point vacuum graph ($f=0$) all have singular order $\omega =0$.
Besides, the $4$-point lepton-photon scattering graphs and
the $4$-point (fermionic) self-energy graphs ($f=2$) are regular distributions with singular order $%
\omega =-3$. Finally, the $4$-point lepton-lepton scattering graphs ($f=4$)
are also regular with singular order $\omega =-6$.

\item The $3$-point graphs have singular order: $\omega _{3}=1-\dfrac{3}{2}f 
$. Hence, the vertex graph with an odd number of non-contracted electromagnetic
graphs ($f=0$) has singular order $\omega =1$. But, from Furry's theorem 
\footnote{Following the same analysis
developed for the causal approach of the QED$_{4}$ in Ref.~\cite{ref21},
it is possible to show that the GQED$_4$ model does not break any discrete symmetries.
 Therefore, in particular, GQED$_4$ satisfies the Furry's theorem.} these graphs do not contribute. Also,
the usual vertex graph ($f=2$) is a regular distribution with singular order 
$\omega =-2$, then, in contrast with QED$_{4}$ results, this contribution is
UV finite, as it was shown in Ref.~\cite{ref15}.

\item The $2$-point graphs have singular order: $\omega _{2}=2-\dfrac{3}{2}
f $. Hence, the one-loop vacuum polarization and the $2$-point vacuum graph (%
$f=0$) have singular order $\omega =2$. The $2$-point lepton-photon
scattering graph and the one-loop self-energy ($f=2$) are regular
distributions with singular order $\omega =-1$. Again, in contrast with QED$%
_{4}$ results, the one-loop self-energy does not have UV divergences, as
shown in Ref.~\cite{ref15}. The $2$-point lepton-lepton scattering graph ($%
f=2$) is also regular, with singular order $\omega =-4$.

\item Finally we can assign to the fundamental vertex the singular order $%
\omega _{1}=0$. This is a practical assumption that allows us to deduce the
singular order of graphs contracted with a fundamental vertex by means of Eq.%
\eqref{Contrafor}.
\end{enumerate}


\section{Gauge invariance}

\label{sec:5}

To study the gauge structure of GQED$_{4}$ in the causal approach, we shall
analyse how the S-matrix series%
\begin{equation}
S\left[ g\right] =1+\sum\limits_{n=1}^{\infty }\frac{1}{n!}\int
dx_{1}dx_{2}...dx_{n}T_{n}\left( x_{1},x_{2},...,x_{n}\right) g\left(
x_{1}\right) g\left( x_{2}\right) ...g\left( x_{n}\right) 
\end{equation}%
behave upon a gauge transformation. Since the causal approach only takes into account
free fields, then we must formulate the gauge transformation upon the electromagnetic
free field $(A_{\mu })$ and the free Dirac fields $(\psi ,\bar{\psi})$. However, the
latter fields are not gauge invariant. Therefore, we only take into account the gauge transformation
on the electromagnetic free field. Furthermore, we shall consider a gauge
transformation by means of a \emph{classical} (c-number) gauge function $\Lambda \left(
x\right) $%
\begin{equation}
A_{\mu }\left( x\right) \rightarrow A_{\mu }\left( x\right) -\frac{1}{e}
\partial _{\mu }\Lambda \left( x\right) ,  \label{GaugeTr1}
\end{equation}%
where $\Lambda \left( x\right) $ vanishes at infinity and, since we consider
the non-mixing gauge fixing condition Eq.\eqref{Non-mix-cond}, this function satisfies
the pseudo-differential equation: $\left( 1+a^{2}\square \right)
^{1/2}\square \Lambda =0$. Our goal here is to show that the S-matrix is invariant under
the gauge transformation \eqref{GaugeTr1}. Hence, we need to calculate how each  perturbative term transforms individually under \eqref{GaugeTr1}.


\subsection{Gauge transformation}

In the same way as we did in our previous analysis, our discussion here will
follow the inductive procedure. Thus, we start by analysing how the first
perturbative term $S_{1}\left[ g\right] =\int dxT_{1}\left( x\right) g\left(
x\right) $ changes under the gauge transformation \eqref{GaugeTr1} 
\begin{align}
S ^{\prime} _{1}\left[ g\right] =& S_{1}\left[ g\right] +i\int dx\partial
_{\mu }\left[ \colon \bar{\psi}\left( x\right) \gamma ^{\mu }\psi \left(
x\right) \colon \right] \Lambda \left( x\right) g\left( x\right)  \notag \\
&+i\int dx\colon \bar{\psi}\left( x\right) \gamma ^{\mu }\psi \left(
x\right) \colon \Lambda \left( x\right) \partial _{\mu }g\left( x\right) .
\end{align}
Now, recalling the current conservation from the Dirac field equation 
\begin{equation}
\partial _{\mu }\left[ \colon \bar{\psi}\left( x\right) \gamma ^{\mu }\psi
\left( x\right) \colon \right] =0,  \label{CurrCon}
\end{equation}%
and if we also consider that $\partial _{\mu }g\left( x\right) =0$, we find
 that the term $S_{1}\left[ g\right] $ is gauge invariant. The last condition is
naturally achieved when the adiabatic limit is considered, $g\rightarrow 1$,
which represents the interacting physical system without external
influences. Therefore, we shall study the gauge transformation of all
perturbative contributions under the adiabatic limit, 
\begin{equation}
S_{n}\left[ 1\right] =\frac{1}{n!}\int dx_{1}dx_{2}...dx_{n}T_{n}\left(
x_{1},x_{2},...,x_{n}\right) .
\end{equation}%
As aforementioned, we will only consider the transformation of the type %
\eqref{GaugeTr1} in this work. Thus, it is convenient to express the $n$%
-point distribution $T_{n}$ in terms of the normally ordered product of
photon operators 
\begin{equation}
T_{n}\left( x_{1},...,x_{n}\right) =\sum\limits_{l=0}^{n}\mathcal{T}%
_{k_{1}\cdots k_{l}}^{\mu _{1}\ldots \mu _{l}}\left( x_{1},...,x_{n}\right)
\colon A_{\mu _{1}}\left( x_{k_{1}}\right) \cdots A_{\mu _{l}}\left(
x_{k_{l}}\right) \colon , \label{distgauge}
\end{equation}%
where $\mathcal{T}_{k_{1}\cdots k_{l}}^{\mu _{1}\ldots \mu _{l}}$ contains
the numerical part and the non-contracted spinor fields. Since $\Lambda $ is
a c-number, we can observe that the electromagnetic contraction is gauge
invariant%
\begin{equation}
\left[ A^{\prime \left( -\right) }_{\mu}\left(
x\right) ,A^{\prime \left( +\right) }_{ \nu}\left( y\right) \right] =\left[ A^{ \left( -\right) }_{\mu}\left( x\right) ,A^{\left( +\right) }_{
\nu} \left( y\right) \right] =iD^{
\left( +\right) }_{\mu \nu}\left( x-y\right) .
\end{equation}%
Then we can immediately conclude that the quantity $\mathcal{T} _{k_{1}\cdots
k_{l}}^{\mu _{1}\ldots \mu _{l}}$ is also gauge invariant. Hence, the
distribution $T_{n}$ \eqref{distgauge} has the following form under gauge transformation
\begin{equation}
T^{\prime} _{n}\left( x_{1},...,x_{n}\right) =\sum\limits_{l=0}^{n}\mathcal{T%
}_{k_{1}\cdots k_{l}}^{\mu _{1}\ldots \mu _{l}}\left( x_{1},...,x_{n}\right)
\colon A^{\prime} _{\mu _{1}}\left( x_{k_{1}}\right) \cdots A^{\prime} _{\mu
_{j}}\left( x_{k_{j}}\right) \cdots A^{\prime} _{\mu _{l}}\left(
x_{k_{l}}\right) \colon .
\end{equation}%
Now applying the gauge transformation \eqref{GaugeTr1} into the field $%
A^{\prime} _{\mu _{j}}$, we show that the transformed contribution $%
S^{\prime} _{n} \equiv S^{\prime} _{n}\left[ 1\right] $ can be written as follows%
\begin{align}
S^{\prime} _{n} &=\frac{1}{n!}\sum\limits_{l=0}^{n}\int dx_{1}...dx_{n}%
\mathcal{T}_{k_{1}\cdots k_{l}}^{\mu _{1}\ldots \mu _{l}}\left(
x_{1},...,x_{n}\right) \colon A^{\prime} _{\mu _{1}}\left( x_{k_{1}}\right)
\cdots A_{\mu _{j}}\left( x_{k_{j}}\right) \cdots A^{\prime} _{\mu
_{l}}\left( x_{k_{l}}\right) \colon  \notag \\
&\quad +\frac{1}{n!}\sum\limits_{l=0}^{n}\frac{1}{e}\int dx_{1}...dx_{n}\left[
\partial _{\mu _{j}}\mathcal{T}_{k_{1}\cdots k_{l}}^{\mu _{1}\ldots \mu
_{l}}\left( x_{1},...,x_{n}\right) \right] \colon A^{\prime} _{\mu
_{1}}\left( x_{k_{1}}\right) \cdots \Lambda \left( x_{k_{j}}\right) \cdots
A^{\prime} _{\mu _{l}}\left( x_{k_{l}}\right) \colon .
\end{align}%
Hence, we see that in order to achieve a gauge invariant theory, we need to analyse whether the
condition%
\begin{equation}
\frac{\partial }{\partial x_{k_{j}}^{\mu _{j}}}\mathcal{T}_{k_{1}\cdots
k_{l}}^{\mu _{1}\ldots \mu _{l}}\left( x_{1},...,x_{n}\right) =0,
\label{CondGAu}
\end{equation}%
is satisfied $\forall ~j$, $\forall ~l$ and $\forall ~n$.

The above condition is now analyzed through the inductive method.
Taking into account the first-order element, $\mathcal{T}_{1}^{\mu }\left( x_{1}\right) =ie\colon \bar{\psi}\left( x\right) \gamma ^{\mu }\psi \left( x\right) \colon $, we easily see that 
\begin{equation}
\partial _{\mu }\mathcal{T}_{1}^{\mu }\left( x_{1}\right) =0,
\end{equation}%
where we have used the result \eqref{CurrCon}. Besides, for higher-order terms we consider the inductive hypothesis, i.e. the condition $\left( \ref{CondGAu}%
\right) $ is valid $\forall ~m\leq n-1$, this means that%
\begin{equation}
\frac{\partial }{\partial x_{k_{j}}^{\mu _{j}}}\mathcal{T}_{k_{1}\cdots
k_{l}}^{\mu _{1}\ldots \mu _{l}}\left( x_{1},...,x_{m}\right) =0, \quad
\forall ~j,~~\forall ~l, ~~\text{and}~~\forall ~m\leq n-1.
\end{equation}%
Thus, from such result follows as well that the $n$-order intermediate distributions valued in the
adiabatic limit
\begin{align}
\int dx_{1}...dx_{n}R_{n}^{\prime }\left( x_{1},...,x_{n}\right)=&\int
dx_{1}...dx_{n}\sum_{P_{2}}T_{n-n_{1}}\left( Y,x_{n}\right) \tilde{T}%
_{n_{1}}\left( X\right) , \\
\int dx_{1}...dx_{n}A_{n}^{\prime }\left( x_{1},...,x_{n}\right) =&\int
dx_{1}...dx_{n}\sum_{P_{2}}\tilde{T}_{n_{1}}\left( X\right)
T_{n-n_{1}}\left( Y,x_{n}\right) ,
\end{align}%
are gauge invariant. Hence, it follows from the definition \eqref{Ddef} that the causal distribution $D_{n}$ valued in the adiabatic limit is gauge invariant as well. Furthermore, if we write $D_{n}$ as 
\begin{equation}
D_{n}\left( x_{1},...,x_{n}\right) =\sum\limits_{l=0}^{n}\mathcal{D}%
_{k_{1}\cdots k_{l}}^{\mu _{1}\ldots \mu _{l}}\left( x_{1},...,x_{n}\right)
\colon A_{\mu _{1}}\left( x_{k_{1}}\right) \cdots A_{\mu _{l}}\left(
x_{k_{l}}\right) \colon ,
\end{equation}%
one can show that 
\begin{equation}
\dfrac{\partial }{\partial x_{k_{j}}^{\mu _{j}}}\mathcal{D}_{k_{1}\cdots
k_{l}}^{\mu _{1}\ldots \mu _{l}}\left( x_{1},...,x_{n}\right) =0, \quad
\forall ~j,\quad \forall ~l.
\end{equation}%
Thus, since $T_{n}$ is defined such as: $T_{n}=R_{n}-R_{n}^{\prime }$,
we need to analyse now how the distribution $R_{n}$ is transformed.
Then, if the retarded distribution $R_{n}$ is written as%
\begin{equation}
R_{n}\left( x_{1},...,x_{n}\right) =\sum\limits_{l=0}^{n}\mathcal{R}%
_{k_{1}\cdots k_{l}}^{\mu _{1}\ldots \mu _{l}}\left( x_{1},...,x_{n}\right)
\colon A_{\mu _{1}}\left( x_{k_{1}}\right) \cdots A_{\mu _{l}}\left(
x_{k_{l}}\right) \colon ,
\end{equation}%
and the condition 
\begin{equation}
\frac{\partial }{\partial x_{k_{j}}^{\mu _{j}}}\mathcal{R}_{k_{1}\cdots
k_{l}}^{\mu _{1}\ldots \mu _{l}}\left( x_{1},...,x_{n}\right) =0, \quad
\forall ~j,\quad \forall ~l,  \label{RGaucond}
\end{equation}%
is satisfied, then the gauge invariance of $R_{n}$ is automatically proved.
Moreover, since $R_{n}$ is the retarded part of $D_{n}$ we have that%
\begin{equation}
\mathcal{R}_{k_{1}\cdots k_{l}}^{\mu _{1}\ldots \mu _{l}}= 
\begin{cases}
\mathcal{D}_{k_{1}\cdots k_{l}}^{\mu _{1}\ldots \mu _{l}} & \text{ on }~~
\Gamma _{n-1}^{+}\left( x_{n}\right) \slash \left\{ \left(
x_{1},...,x_{n}\right) \right\} \\ 
0 & \text{ on} ~~ \left( \Gamma _{n-1}^{+}\left( x_{n}\right) \right) ^{C},%
\end{cases}%
\end{equation}
where $ C $ denotes the complement. From this result we find that the condition \eqref{RGaucond} is almost
satisfied 
\begin{equation}
\frac{\partial }{\partial x_{k_{j}}^{\mu _{j}}}\mathcal{R}_{k_{1}\cdots
k_{l}}^{\mu _{1}\ldots \mu _{l}}\left( x_{1},...,x_{n}\right) =0, \quad 
\text{on }\quad \left(  \Gamma _{n-1}^{+}\left( x_{n}\right) \slash 
\left\{ \left( x_{1},...,x_{n}\right) \right\} \right) \cup \left( \Gamma
_{n-1}^{+}\left( x_{n}\right) \right) ^{C}.
\end{equation}%
Thus, the distribution $\frac{\partial }{\partial x_{k_{j}}^{\mu _{j}}} 
\mathcal{R}_{k_{1}\cdots k_{l}}^{\mu _{1}\ldots \mu _{l}}$ has causal
support at the origin. Furthermore, since $R_{n}^{\prime }$ is gauge invariant,
and by the property $T_{n}=R_{n}-R_{n}^{\prime }$, we can conclude that the
distribution $\frac{\partial }{\partial x_{k_{j}}^{\mu _{j}}}\mathcal{T}%
_{k_{1}\cdots k_{l}}^{\mu _{1}\ldots \mu _{l}}$ has also causal support at
the origin.

It is interesting to remark that we can write $\mathcal{T}_{k_{1}\cdots
k_{l}}^{\mu _{1}\ldots \mu _{l}}$ as the contribution of all possible
configuration of non-contracted spinor fields%
\begin{equation}
\mathcal{T}_{k_{1}\cdots k_{l}}^{\mu _{1}\ldots \mu _{l}}\left(
x_{1},...,x_{n}\right) =\sum\limits_{f=0}^{2n}\left[ ^{f}\mathcal{T}%
_{k_{1}\cdots k_{l}}^{\mu _{1}\ldots \mu _{l}}\left( x_{1},...,x_{n}\right) %
\right] ,
\end{equation}%
where $^{f}\mathcal{T}_{k_{1}\cdots k_{l}}^{\mu _{1}\ldots \mu _{l}}$
contains the numerical part and the non-contracted spinor fields 
\begin{equation}
^{f}\mathcal{T}_{k_{1}\cdots k_{l}}^{\mu _{1}\ldots \mu _{l}}\left(
x_{1},...,x_{n}\right) \equiv \text{ }\sum_{g_{l}}\colon
\prod\limits_{j=1}^{f_{g}}\bar{\psi}\left( x_{i_{j}}\right) \left[ ^{\mu
_{m_{1}}\cdots \mu _{m_{l}}}t_{n_{1}\cdots n_{f_{g}}}^{i_{1}\cdots
i_{f_{g}}}\left( x_{1},x_{2},...,x_{n}\right) \right] \prod%
\limits_{j=1}^{f_{g}}\psi \left( x_{n_{j}}\right) \colon ,
\end{equation}%
where the index $\left\{ k_{1}\cdots k_{l}\right\} $ are implicitly on the
right-hand side of the above expression. Finally, since each class of graphs
is independent, we have that each one of these contributions is also individually
independent. Therefore, each contribution has causal support at the origin and by a theorem
\footnote{%
If the support of a distribution $f\in D^{\prime }\left( \supset \mathcal{J}%
^{\prime }\right) $ contains a single point $x=0$, then it is unambiguously
represented in the form: $f\left( x\right) =\sum\limits_{b=0}^{\omega
}c_{b}D^{b}\delta \left( x\right) $, where $\omega $ is the singular order
of $f$. This holds equivalently in the Fourier transformed space: $\hat{f}%
\left( p\right) =\sum\limits_{b=0}^{\omega }c_{b}p^{b}$. For further detail
see \cite{ref36}.} they have the form 
\begin{equation}
\dfrac{\partial }{\partial x_{k_{j}}^{\mu _{j}}}\left[ ^{f}\mathcal{T}
_{k_{1}\cdots k_{l}}^{\mu _{1}\ldots \mu _{l}}\right] =\sum\limits_{g_{f,l}}%
\colon \prod\limits_{j=1}^{f}\bar{\psi}\left( x_{i_{j}}\right) \left[
\sum\limits_{\left\vert a\right\vert =0}^{\omega \left( g_{f,l}\right)
+1}C_{a}^{g_{f,l}}D^{a}\delta \left( x_{1}-x_{n}\right) \cdots \delta \left(
x_{n-1}-x_{n}\right) \right] \prod\limits_{j=1}^{f}\psi \left(
x_{n_{j}}\right) \colon ,  \label{GaugeCond}
\end{equation}%
where the summation is over all possible graphs with $2f$ and $l$
non-contracted spinor and electromagnetic fields, respectively. In particular, it is considered 
that each graph $g_{f,l}$ has singular order $\omega \left( g_{f,l}\right) $.

Finally, the analysis of the gauge structure of GQED$_{4}$ was reduced to
the study of the expression \eqref{GaugeCond}. Moreover, one can easily see that only graphs 
with singular order $\omega \left( g_{f,l}\right) \geq -1$
contribute. We can therefore conclude that from the results of the
previous section, only graphs with less than $6$-point shall contribute for the analysis. 
Each nonzero case is usually called an anomaly, because it violates
gauge invariance. Now, we analyse individually each possible case:

\begin{enumerate}
\item The $5$-point graphs have singular order: $\omega _{5}=-1-\dfrac{3}{2}%
f $. Hence, graphs containing only non-contracted electromagnetic fields ($f=0$) 
and with singular order $\omega _{5}=-1$ may be an anomaly. However, by Furry's theorem, these graphs do not contribute.

\item The $4$-point graphs have singular order: $\omega _{4}=-\dfrac{3}{2}f$%
. Hence, the light-light scattering and the vacuum polarization graphs,\footnote{The $4$-point
 vacuum graphs do not have external electromagnetic fields, thus, it is not considered.} each one with $f=0$, have singular order $%
\omega =0$. We analyse each case separately:

In the vacuum graph all electromagnetic field are contracted; besides, since
this type of contraction is gauge invariant, this graph is too.

We can assume that the $4$-point vacuum polarization graphs have the
non-contracted electromagnetic fields in $x_{1}$ and $x_{4}$, we have that 
\begin{equation}
^{0}\mathcal{T}_{VP}^{\mu _{1}\mu _{4}}\left( x_{1},...,x_{4}\right)
=\sum_{g_{0,2}}t^{\mu _{1}\mu _{4}}\left( x_{1},x_{2},x_{3},x_{4}\right)
\equiv \Pi ^{\mu _{1}\mu _{4}}\left( x_{1},x_{2},x_{3},x_{4}\right) ,
\end{equation}%
is a numerical distribution. Furthermore, from Eq.\eqref{GaugeCond}, we find that the derivative with respect to $x_{1}$ is given by%
\begin{equation}
\partial _{\mu _{1}}\left[ \Pi ^{\mu _{1}\mu _{4}}\left(
x_{1},...,x_{4}\right) \right] =\sum_{k=1}^{3}C_{4k}\frac{\partial }{%
\partial \left( x_{k}\right) _{\mu _{4}}}\left[ \delta \left(
x_{1}-x_{4}\right) \cdots \delta \left( x_{3}-x_{4}\right) \right] ,
\end{equation}%
where $\partial ^{\mu _{j}}\equiv \frac{\partial }{%
\partial \left( x_{j}\right) _{\mu _{j}}}$. We also omit the constant term
because it is not Lorentz invariant. Moreover, since $\partial _{\mu
_{1}}\partial _{\mu _{4}}\left[ \Pi ^{\mu _{1}\mu _{4}}\right] $ is
invariant with respect to the exchange $x_{1}\leftrightarrow x_{4}$, we obtain
\begin{equation}
\partial _{\mu _{1}}\left[ \Pi ^{\mu _{1}\mu _{4}}\left(
x_{1},...,x_{4}\right) \right] =\partial _{\mu _{1}}\left[ Cg^{\mu _{1}\mu
_{4}}\delta \left( x_{1}-x_{4}\right) \cdots \delta \left(
x_{3}-x_{4}\right) \right] ,
\end{equation}%
a similar relation can be obtained for the derivative with respect to $x_{4}$%
. Thus, we conclude that the quantity between the square bracket is a distribution
of singular order $\omega \left( \Pi ^{\mu _{1}\mu _{4}}\right) =0$. Hence,
the initial impression of an anomaly can be removed by a normalization of
$\Pi ^{\mu _{1}\mu _{4}}$. Finally, we show that the $4$-point vacuum 
polarization is a gauge invariant distribution and a transversal tensor 
\begin{equation}
\partial _{\mu _{j}}\Pi ^{\mu _{1}\mu _{4}}\left(
x_{1},x_{2},x_{3},x_{4}\right) =0,\quad j=1,4.
\end{equation}

For the $4$-point light-light scattering graph, which has all the spinor
field contracted, we have that the quantity%
\begin{equation}
^{0}\mathcal{T}_{LL}^{\mu _{1}\ldots \mu _{4}}\left( x_{1},...,x_{4}\right)
=t^{\mu _{1}\cdots \mu _{4}}\left( x_{1},x_{2},x_{3},x_{4}\right) ,
\end{equation}%
is a numerical distribution. Hence, from Eq.\eqref{GaugeCond}, we have that
its derivative with respect to $x_{1}$ is given by%
\begin{align}
\partial _{\mu _{1}}\left[ ^{0}\mathcal{T}_{LL}^{\mu _{1}\ldots \mu
_{4}}\left( x_{1},...,x_{4}\right) \right] &= \left[ \sum_{k=1}^{3}\left(
C_{2k}\partial _{k}^{\mu _{2}}g^{\mu _{3}\mu _{4}}+C_{3k}\partial _{k}^{\mu
_{3}}g^{\mu _{2}\mu _{4}}+C_{4k}\partial _{k}^{\mu _{4}}g^{\mu _{2}\mu
_{3}}\right) \right]  \notag \\
& \quad \times \delta \left( x_{1}-x_{4}\right) \cdots \delta \left(
x_{3}-x_{4}\right) .
\end{align}%
Since $\partial _{\mu _{1}}\partial _{\mu _{2}}\partial _{\mu _{3}}\partial
_{\mu _{4}}\left[ ^{0}\mathcal{T}_{LL}^{\mu _{1}\ldots \mu _{4}}\right] $ is
symmetric with respect to the permutation of the derivative variables, we
have that
\begin{align}
\partial _{\mu _{1}}\left[ ^{0}\mathcal{T}_{LL}^{\mu _{1}\ldots \mu
_{4}}\left( x_{1},...,x_{4}\right) \right] &= \partial _{\mu _{1}}\left[
C\left( g^{\mu _{1}\mu _{2}}g^{\mu _{3}\mu _{4}}+g^{\mu _{1}\mu _{3}}g^{\mu
_{2}\mu _{4}}+g^{\mu _{1}\mu _{4}}g^{\mu _{2}\mu _{3}}\right) \right]  \notag \\
& \quad \times \delta \left( x_{1}-x_{4}\right) \cdots \delta \left(
x_{3}-x_{4}\right) .
\end{align}%
A similar relation can be obtained for the other derivative variables. 
Like in the previous case, the anomaly can be removed by
a normalization of $^{0}\mathcal{T}_{LL}^{\mu _{1}\ldots \mu _{4}}$.
Therefore, the $4$-point light-light scattering graph is a gauge invariant
distribution and transversal tensor as well%
\begin{equation}
\partial _{\mu _{j}}t^{\mu _{1}\cdots \mu _{4}}\left(
x_{1},x_{2},x_{3},x_{4}\right) =0,\quad j=1,2,3,4.
\end{equation}

\item The $3$-point graphs have singular order: $\omega _{3}=1-\dfrac{3}{2}f$%
. Hence, the only possible anomaly are those vertex graphs with three
and one non-contracted electromagnetic graphs ($f=0$), both graphs have singular 
order $\omega =1$; but, according to Furry's theorem these graphs do not contribute.

\item The $2$-point graphs have singular order: $\omega _{2}=2-\dfrac{3}{2}f$%
. However, the $2$-point vacuum graph $\left( \omega =2\right) $, the one-loop
self-energy $\left( \omega =-1\right) $ and the lepton-lepton scattering
graph $\left( \omega =-1\right) $ do not have external
electromagnetic fields. Besides, since this type of contraction is gauge
invariant, these graphs are gauge invariant as well.

The one-loop vacuum polarization has singular order $\omega =2$, and
following the procedure as in Ref.~\cite{ref21}, it can be transformed by a
normalization into a gauge invariant distribution.

The $2$-point lepton-photon scattering graph is a regular
distribution with singular order $\omega =-1$. Since this graph has two
contributions, with help of \eqref{GaugeCond}, we have that 
\begin{align}
\partial _{\mu _{1}}\left[ ^{2}\mathcal{T}_{Com}^{\mu _{1}\mu _{2}}\left(
x_{1},x_{2}\right) \right] &=C_{12}\colon \bar{\psi}\left( x_{1}\right)
\delta \left( x_{1}-x_{2}\right) \gamma ^{\mu _{2}}\psi \left( x_{2}\right)
\colon \nonumber \\
&\quad +C_{21}\colon \bar{\psi}\left( x_{2}\right) \delta \left(
x_{2}-x_{1}\right) \gamma ^{\mu _{2}}\psi \left( x_{1}\right) \colon .
\end{align}
Further, using the Dirac equation for the fermionic propagator $S^{F}$, the above
formula can be transformed as follows 
\begin{align}
\partial _{\mu _{1}}\left[ ^{2}\mathcal{T}_{Com}^{\mu _{1}\mu _{2}}\left(
x_{1},x_{2}\right) \right] &=i\partial _{\mu _{1}}\Big[ -C_{12}\colon \bar{%
\psi}\left( x_{1}\right) \gamma ^{\mu _{1}}S^{F}\left( x_{1}-x_{2}\right)
\gamma ^{\mu _{2}}\psi \left( x_{2}\right) \colon    \notag \\
&\quad   +C_{21}\colon \bar{\psi}\left( x_{2}\right) \gamma ^{\mu
_{2}}S^{F}\left( x_{2}-x_{1}\right) \gamma ^{\mu _{1}}\psi \left(
x_{1}\right) \colon \Big] ,
\end{align}%
the quantity on the square bracket is a distribution of singular order $\omega
\left( ^{2}\mathcal{T}_{Com}^{\mu _{1}\mu _{2}}\right) =-1$. If we recall to 
the invariance under charge conjugation, we have that $C_{12}=-C_{21}$. Hence,
since the quantity $\partial _{\mu _{1}}\left[ ^{2}\mathcal{T}_{Com}^{\mu _{1}
\mu _{2}}\left( x_{1},x_{2}\right) \right] $ has support on $x_{1}=x_{2}$, 
we can conclude that this anomaly can be removed. One may obtain
the same result when considering the derivative with respect to $x_{2}$.
Finally, we find that 
\begin{equation}
\partial _{\mu _{j}}\left[ ^{2}\mathcal{T}_{Com}^{\mu _{1}\mu _{2}}\left(
x_{1},x_{2}\right) \right] =0,\quad j=1,2,
\end{equation}%
which means that the lepton-photon scattering graphs are gauge invariant too.
\end{enumerate}


\subsection{Ward-Takahashi-Fradkin identities}

The Ward-Takahashi-Fradkin (WTF) identities are usually derived in QED$_{4}$
scattering processes when free photons are involved \cite{ref39}. In the framework of
functional method, these (coupled) identities are satisfied by the complete
Green's functions. In contrast with this non-perturbative method, the causal
approach determines these relations perturbatively order-by-order. Using the results 
obtained previously, we will see that the WTF identities for GQED$_{4}$ are easily derived.

\paragraph{Vacuum polarization transversality} The contribution to the vacuum polarization is
 a class of graphs with two non-contracted electromagnetic fields. The transversality 
 condition must be proved for each graph of this class. Hence, for the $n$-order vacuum
  polarization graph we shall assume that the non-contracted electromagnetic fields are at the points $\left( x_{1},x_{2}\right) $, then we have that%
\begin{equation}
^{0}\mathcal{T}_{VP}^{\mu _{1}\mu _{2}}\left( x_{1},x_{2},\ldots
,x_{n}\right) =\Pi ^{\mu _{1}\mu _{2}}\left( x_{1},x_{2},\ldots
,x_{n}\right) ,
\end{equation}%
is a numerical distribution. Moreover, by taking into account Eq.%
\eqref{GaugeCond}, we obtain
\begin{equation}
\partial _{\mu _{j}}\Pi ^{\mu _{1}\mu _{2}}\left( x_{1},x_{2},\ldots
,x_{n}\right) =0,\quad j=1,2,\quad \forall ~n>4.
\end{equation}%
Besides, by a normalization, we have proved this relation for $n=2$
and $n=4$; therefore, the transversality of the vacuum polarization tensor
is guaranteed 
\begin{equation}
\partial _{\mu _{j}}\Pi ^{\mu _{1}\mu _{2}}\left( x_{1},x_{2},\ldots
,x_{n}\right) =0,\quad j=1,2,\quad \forall ~n\geq 2,  \label{eq2}
\end{equation}%
and, by relabelling the derivative variable we can show the validity of this
relation for all the points $\left( x_{3},x_{4},...\right) $.

\paragraph{Vertex and self-energy} For a $n$-point vertex graph, we can choose
the non-contracted electromagnetic field in $x_{n}$ and the two
non-contracted spinor fields in $\left( x_{i},x_{j}\right) $. In general, we
can recognize three different possible graphs: \textit{(i)} when all
non-contracted fields are in different points; \textit{(ii)} when one of the
non-contracted spinor is in the same point of the non-contracted
electromagnetic field, so this vertex is connected to the graph by a spinor
line; \textit{(iii)} when the two non-contracted spinor fields are in the
same point, then this vertex is connected to the graph by an electromagnetic
line. Taking all these cases into account we have that 
\begin{align}
^{2}\mathcal{T}_{Ver}^{\mu _{n}}\left( x_{1},x_{2},\ldots ,x_{n}\right)
&=\sum_{i\neq j}^{n-1}\colon \bar{\psi}\left( x_{i}\right) \Lambda ^{\mu
_{n}}\left( x_{n},x_{i},x_{j},\ldots \right) \psi \left( x_{j}\right) \colon \notag \\
&\quad +\sum_{i\neq j}^{n-1}\colon \bar{\psi}\left( x_{n}\right) \gamma
^{\mu_{n}}S^{F}\left( x_{n}-x_{i}\right) \Sigma \left( x_{i},x_{j},\ldots
\right) \psi \left( x_{j}\right) \colon  \notag \\
&\quad +\sum_{i\neq j}^{n-1}\colon \bar{\psi}\left( x_{i}\right) \Sigma \left(
x_{i},x_{j},\ldots \right) S^{F}\left( x_{j}-x_{n}\right) \gamma ^{\mu
_{n}}\psi \left( x_{n}\right) \colon  \notag \\
&\quad +\sum_{i\neq j}^{n-1}\colon \bar{\psi}\left( x_{i}\right) \Pi ^{\mu_{n}\mu
_{j}}\left( x_{n},x_{j},\ldots ,\right) D_{\mu _{j}\mu _{i}}^{F}\left(
x_{j}-x_{i}\right) \gamma ^{\mu _{i}}\psi \left(x_{i}\right) \colon .
\end{align}
where $\Lambda ^{\mu _{n}}$ and $\Sigma $ are the vertex function and the
self-energy, respectively. Since the vertex function has singular order $-2$%
, it follows from Eq.\eqref{GaugeCond} that 
\begin{equation}
\partial _{\mu _{n}}\left[ ^{2}\mathcal{T}_{Ver}^{\mu
_{n}}\left(x_{1},x_{2},\ldots ,x_{n}\right) \right] =0,  \label{eq3}
\end{equation}%
without requiring any assumption. Moreover, from the transversality
condition for the vacuum polarization \eqref{eq2} and from the following
identities 
\begin{align}
\frac{\partial }{\partial x_{n}^{\mu _{n}}}\left[ \bar{\psi}\left(
x_{n}\right) \gamma ^{\mu _{n}}S^{F}\left( x_{n}-x_{i}\right) \right] &=i 
\bar{\psi}\left( x_{n}\right) \delta \left( x_{n}-x_{i}\right), \\
\frac{\partial }{\partial x_{n}^{\mu _{n}}}\left[ S^{F}\left(x_{j}-x_{n}%
\right) \gamma ^{\mu _{n}}\psi \left( x_{n}\right) \right] &=-i\delta \left(
x_{j}-x_{n}\right) \psi \left( x_{n}\right) ,
\end{align}
we obtain the relation
\begin{align}
\partial _{\mu _{n}}\left[ ^{2}\mathcal{T}_{Ver}^{\mu
_{n}}\left(x_{1},x_{2},\ldots ,x_{n}\right) \right] &=\sum_{i\neq
j}^{n-1}\colon \bar{\psi}\left( x_{i}\right) \partial _{\mu _{n}}\Lambda
^{\mu _{n}}\left( x_{n},x_{i},x_{j},\ldots \right) \psi \left( x_{j}\right)
\colon  \notag \\
&\quad +i\sum_{i\neq j}^{n-1}\colon \bar{\psi}\left( x_{n}\right) \delta \left(
x_{n}-x_{i}\right) \Sigma \left( x_{i},x_{j},\ldots \right) \psi \left(
x_{j}\right) \colon  \notag \\
&\quad -i\sum_{i\neq j}^{n-1}\colon \bar{\psi}\left( x_{i}\right) \Sigma \left(
x_{i},x_{j},\ldots \right) \delta \left( x_{j}-x_{n}\right) \psi \left(
x_{n}\right) \colon .
\end{align}
Finally, using the condition \eqref{eq3} into the above formula we obtain the closed relation
\begin{equation}
\partial _{\mu _{n}}\Lambda ^{\mu _{n}}\left( x_{n},x_{i},x_{j},\ldots
\right) =i\left[ \delta \left( x_{j}-x_{n}\right) \Sigma \left(
x_{i},x_{j},\ldots \right) -\delta \left( x_{n}-x_{i}\right) \Sigma \left(
x_{i},x_{j},\ldots \right) \right] .
\end{equation}%
Nonetheless, since $\Sigma $ and $\Lambda ^{\mu _{n}}$ are translational
invariant, we can rewrite $\Lambda ^{\mu _{n}}$ equivalently now with respect to $x_{n}$ and $%
\Sigma $ with respect to $x_{j}$. Hence, we arrive into the following
Ward-Takahashi-Fradkin identity written in the configuration space 
\begin{equation}
\partial _{\mu _{n}}\Lambda ^{\mu _{n}}\left( x_{i}-x_{n},x_{j}-x_{n},\ldots
\right) =i\left[ \delta \left( x_{j}-x_{n}\right) \Sigma
\left(x_{i}-x_{j},\ldots \right) -\delta \left( x_{i}-x_{n}\right) \Sigma
\left( x_{i}-x_{j},\ldots \right) \right] .
\end{equation}%
Now, considering the change of variable $y_{k}=x_{k}-x_{n}$%
\begin{equation}
-\sum _{k}\partial ^{\nu }_{y_{k}}\Lambda _{\nu }\left( y_{i},y_{j},\ldots
\right) =i\left[ \delta \left( y_{j}\right) \Sigma \left( y_{i}-y_{j},\ldots
\right) -\delta \left( y_{i}\right) \Sigma \left( y_{i}-y_{j},\ldots \right) %
\right] ,
\end{equation}%
and taking the Fourier transform of this last relation, we obtain the desired form for the 
Ward-Takahashi-Fradkin identity written in the momentum space 
\begin{align}
\left( 2\pi \right) ^{2}\left( \sum\limits_{k=1}^{n-1}p_{k }^{\nu}\right) \hat{\Lambda}_{\nu }\left(
p_{i},p_{j},\ldots \right) &=  \hat{\Sigma}%
\left( p_{i},\cdots ,p_{j-1},p_{j+1},\ldots \right) \nonumber \\
&\quad -\hat{\Sigma}\left(
-\left( \sum\limits_{k\neq i}p_{k}\right) ,\cdots ,p_{j-1},p_{j+1},\ldots
\right) .
\end{align}
In particular, let us consider the case $n=3$, $i=1$ and $j=2$, 
\begin{equation}
\left( p_{1}+p_{2}\right) _{\nu }\hat{\Lambda}^{\nu }\left(
p_{1},p_{2}\right) =\left( 2\pi \right) ^{-2}\left[ \hat{\Sigma}\left(
p_{1}\right) -\hat{\Sigma}\left( -p_{2}\right) \right] .
\end{equation}%
Besides, relabelling the momenta such as: $p_{1}\rightarrow p$ and $p_{2}\rightarrow -q$%
, then $\hat{\Lambda}^{\nu }\left( p,-q\right) \rightarrow \hat{\Lambda}%
^{\nu }\left( p,q\right) $, we arrive into the well-known form of the one-loop
Ward-Takahashi-Fradkin identity 
\begin{equation}
\left( p-q\right) _{\nu }\hat{\Lambda}^{\nu }\left( p,q\right) =\left( 2\pi
\right) ^{-2}\left[ \hat{\Sigma}\left( p\right) -\hat{\Sigma}\left( q\right) %
\right] .
\end{equation}%
Finally, taking the on-shell limit $p\rightarrow q$, we find the following relation 
\begin{equation}
\hat{\Lambda}^{\nu }\left( p,p\right) =\left( 2\pi \right) ^{-2}\frac{%
\partial }{\partial p_{\nu }}\hat{\Sigma}\left( p\right) .
\end{equation}%
This relation may be used to find the one-loop vertex (at zero transferred
momentum) from the one-loop self-energy expression.


\section{Conclusion}

\label{sec:6}

In this paper we have discussed the normalizability and gauge invariance of GQED$_{4}$ 
theory in the context of causal perturbation theory, as proposed by Epstein and Glaser. 
These general physical properties are not considered as axioms in this formalism, but their 
validity has been showed via an inductive procedure provided by the Epstein-Glaser causal method.

First, we remark that if the normalizability of GQED$_{4}$ were analysed following the usual 
approach, since the ('bare') fundamental vertex of QED$_{4}$ and GQED$_{4}$ are the same then
 we would naively conclude that these models must have the same (perturbative) normalizability nature.
However, when the singular order of a (complete) graph is properly analysed
by the causal approach, the singular order of the internal lines
(propagators) is taken into account and plays a crucial part in the analysis. Thus, an
 analysis of graphs with internal photon lines showed that they must have lower singular order than their QED$_{4}$ counterparts. 
The singular order formula \eqref{SIngOrdFor}, obtained for any graph, indicates that
 only graphs with less than 5-point may be UV divergent. Therefore, we can conclude that GQED$_{4}$ is a super-normalizable theory.

Furthermore, the general expression for the singular order allowed us to prove that 
the GQED$_{4}$ is almost an explicitly gauge invariant theory. This is because only 
graphs with no more than 5-point may violate gauge invariance. We have proved in 
detail that all these possible anomalies can suitably be removed by a normalization.
 In particular, the transversality property is satisfied for the cases: the two- and 
 four-point vacuum polarization graphs, the $4$-point light-light scattering graph.
  This strong result led us to prove straightforwardly the Ward-Takahashi-Fradkin 
  identities. Both the vacuum polarization transversality and the relation between 
  the vertex and the self-energy are trivially proven within this framework.

In conclusion, we have proved that the generalized quantum electrodynamics is almost 
an ultraviolet finite field theory, and divergences appear only at lower pertubative 
order. Thus, for futures works we intend to compute, via the causal approach, the
 one-loop radiative correction functions: vacuum polarization, self-energy and 
 vertex function, and find explicitly their singular order which must be in accordance with the result obtained in section~\ref{sec:4}.

\subsection*{Acknowledgments}

R.B. thankfully acknowledges CNPq for partial support, Project No. 304241/2016-4 B.M.P. thanks CNPq and CAPES for
partial support and D.E.S. thanks CNPq for partial support.


\appendix


\section{Degree of divergence of GQED$_{4}$ graphs}

\label{Sec:A}

Let us now discuss further methods to determine the degree of divergence of a graph; 
with particular interest in each method prediction that can be used as an approach for
 comparison of accuracy. For this purpose, we consider first the following interaction term for GQED$_{4}$
\begin{equation}
\mathcal{L}_{int}=e:\bar{\psi}\left( x\right) \gamma ^{\mu }\psi
\left(x\right) :A_{\mu }\left( x\right) ,
\end{equation}%
in this vertex we have that the number of fermionic lines is $n_{F}=2$, and the number
of bosonic lines is $n_{L}=1$. The general definition of the superficial degree of
divergence $w$, of an arbitrary connected one-particle irreducible Feynman diagram
$\Gamma $, is the actual degree of divergence of the integration over the region 
of momentum space in which the momenta of all internal lines go to infinity together.
This is the number of factors of momentum in the numerator minus the number in the 
denominator of the integrand, plus four for every independent four-momentum over which we integrate.

To calculate $w$, we will need to know the following detail about the diagram:%
\begin{align*}
I_{F} &\equiv \text{number of internal fermionic lines,} \\
I_{L} &\equiv \text{number of internal electromagnetic lines,} \\
f &\equiv \text{number of external fermionic lines,} \\
l &\equiv \text{number of external electromagnetic lines,} \\
n &\equiv \text{number of vertices of interaction.}
\end{align*}%
The asymptotic behaviour of the fermionic free Feynman propagator is given by%
\begin{equation}
\hat{S}^{F}\left( p\right) =\left( 2\pi \right) ^{-2}\frac{\left( \gamma
.p+m\right) }{p^{2}-m^{2}++i0^{+}}\sim \left( \gamma .p\right) ^{-1},
\end{equation}%
thus, it has the power momentum 
\begin{equation}
w_{F}=-1,  \label{DivF}
\end{equation}%
as the usual. However, for the Bopp-Podolsky electromagnetic free Feynman
propagator $\left( \xi =1\right) $:%
\begin{equation}
\hat{D}_{\mu \nu }^{F}\left( k\right) =-\left( 2\pi \right) ^{-2}g_{\mu \nu
}\left( \frac{1}{k^{2}+i0^{+}}-\frac{1}{k^{2}-m_{a}^{2}+i0^{+}}\right) \sim
m_{a}^{2}k^{-4},
\end{equation}%
we have the power momentum 
\begin{equation}
w_{L}=-4.  \label{DivL}
\end{equation}%
Adding the contributions from the propagators and total number of
independent momentum variables of integration, we have that
\begin{align}
w\left( \Gamma _{n}\right) =&I_{F}w_{F}+I_{L}w_{L}+4\left[
I_{F}+I_{L}-\left( n-1\right) \right] , \\
=&4+I_{F}\left( w_{F}+4\right) +I_{L}\left( w_{L}+4\right) -4n,
\end{align}%
using the topological identities $2I_{L}+l=n$ and $2I_{F}+f=2n$, we obtain
that%
\begin{equation}
w\left( \Gamma _{n}\right) =4-\left( f\frac{\left( w_{F}+4\right) }{2}+l%
\frac{\left( w_{L}+4\right) }{2}\right) +n\left( w_{F} + \frac{\left(
w_{L}+4\right) }{2}\right) \text{.}  \label{Degre0}
\end{equation}%
Finally, replacing \eqref{DivF} and \eqref{DivL} into \eqref{Degre0}, 
we find that the degree of divergence of a $n$-point GQED$_{4}$ connected graph is given by
\begin{equation}
w\left( \Gamma _{n}\right) =4-\frac{3}{2}f-n\text{,}  \label{Degre}
\end{equation}%
this result is identical to the one obtained by the causal approach %
\eqref{SIngOrdFor}.

Nonetheless, the degree of divergence of a connected graph can also be obtained 
by a simple dimensional analysis. \footnote{This method has the advantage that 
it does not consider the structure of Feynman diagrams.} Considering that 
the action must be dimensionless in natural units ($\hbar =c =1$), so each 
term in the Lagrangian density must have length dimensionality $+4$,%
\begin{equation}
\mathcal{L}_{GQED}=\mathcal{L}_{D}+\mathcal{L}_{P}+\mathcal{L}_{int}.
\end{equation}%
Then, for the fermionic part, $\mathcal{L}_{D}=\bar{\psi}\left( i\gamma
.\partial -m\right) \psi $, it follows that
\begin{equation}
\left[ \psi \right] =\left[ \bar{\psi}\right] =3/2.
\end{equation}
Besides, from the Bopp-Podolsky electromagnetic theory%
\begin{equation}
\mathcal{L}_{P}=-\dfrac{1}{4}F_{\mu \nu }F^{\mu \nu }+\dfrac{a^{2}}{2}%
\partial _{\mu }F^{\mu \sigma }\partial ^{\nu }F_{\nu \sigma }-\dfrac{1}{%
2\xi }\left( \partial .A\right) \left( 1+a^{2}\square \right) \left(
\partial .A\right) ,
\end{equation}%
we have that 
\begin{equation}
\left[ A_{\mu }\right] =1,
\end{equation}%
and also $\left[ \xi \right] =0$ and $\left[ a\right] =-1$. Moreover, from
the interaction part we obtain that the dimensionality for the coupling constant reads
\begin{equation}
\left[ e\right] =4-2\left[ \psi \right] -\left[ A_{\mu }\right] .
\label{DimCar}
\end{equation}%
In general, the free propagator of a field is a four-dimensional Fourier
transform of the vacuum expectation value of a time-ordered product of a 
pair of those free fields. For instance, in GQED$_{4}$ we have the two propagators
\begin{align}
\left( 2\pi \right) ^{-2}\int d^{4}x\left\langle 0\right\vert T\left\{ \psi
_{a}\left( x\right) ,\bar{\psi}_{b }\left( 0\right) \right\} \left\vert
0\right\rangle e^{ipx} =&\hat{S}_{ab}^{F}\left( p\right), \\
\left( 2\pi \right) ^{-2}\int d^{4}x\left\langle 0\right\vert T\left\{
A_{\mu }\left( x\right) ,A_{\nu }\left( 0\right) \right\} \left\vert
0\right\rangle e^{ikx} =&\hat{D}_{\mu \nu }^{F}\left( k\right) .
\end{align}%
From these very definitions we obtain the following relations%
\begin{align}
\left[ \psi \right] =\frac{\left[ \hat{S}^{F}\right] +4}{2}, \quad \left[
A_{\mu }\right] =\frac{\left[ \hat{D}_{\mu \nu }^{F}\right] +4}{2}.
\label{DimA}
\end{align}%
Replacing these results back into \eqref{DimCar} we have that%
\begin{equation}
\left[ e\right] =4-\left( \left[ \hat{S}^{F}\right] +4\right) -\frac{\left( %
\left[ \hat{D}_{\mu \nu }^{F}\right] +4\right) }{2}.  \label{DimE}
\end{equation}%
Applying the same dimensional analysis for the time-ordered part of the $n+f+l$ 
interaction Lagrangian formed by a $n$-point connected Feynman graph
 $\Gamma _{n}$ with $f$ external fermionic lines and $l$ external 
 electromagnetic lines, contracted with $f+l$ vertices, we find that%
\begin{equation}
\left[ \Gamma _{n}\right] =4-f\left( \left[ \hat{S}^{F}\right] +4-\left[
\psi \right] \right) -l\left( \left[ \hat{D}_{\mu \nu }^{F}\right] +4-\left[
A_{\mu }\right] \right) -n\left[ e\right].
\end{equation}%
Furthermore, using the results \eqref{DimA} it follows
\begin{equation}
\left[ \Gamma _{n}\right] =4-f\left[ \psi \right]
 -l\left[ A_{\mu }\right] -n\left[ e\right] .  \label{DimGam}
\end{equation}%
Finally, making use of previous results $\left[ \psi \right] =3/2$, 
$\left[ A_{\mu }\right] =1$ and $\left[ e%
\right] =0$, we find that the degree of divergence obtained
 from a dimensional analysis is
\begin{equation}
\left[ \Gamma _{n}\right] =4-\frac{3}{2}f-l.
\end{equation}%
In the usual approach it is assumed that $\left[ \Gamma _{n}\right]
=w\left(\Gamma _{n}\right) $, but we see that this clearly contradicts 
the previous result \eqref{Degre}. However, if we take an alternative route on the
 analysis, by replacing instead \eqref{DimA} and \eqref{DimE} 
 into \eqref{DimGam} we arrive into 
\begin{equation}
\left[ \Gamma _{n}\right] =4-\left( f\frac{\left( \left[ \hat{S}^{F}\right]
 +4\right) }{2}-l\frac{\left( \left[ \hat{D}_{\mu \nu }^{F}\right] +4\right) 
  }{2}\right) -n\left( 4-\left( \left[ \hat{S}^{F}\right] +4\right) 
  -\frac{\left( \left[ \hat{D}_{\mu \nu }^{F}\right] +4\right) }{2}\right) .
\end{equation}%
We see that this formula is almost similar to \eqref{Degre0}.
Actually, for QED$_{4}$ they are equivalent, but for GQED$_{4}$ the central problem is given by the fact that%
\begin{equation}
\left[ \hat{D}_{\mu \nu }^{F}\right] =-2\neq w_{L} =-4 .
\end{equation}


\section{Distributional Fourier transform}

\label{Sec:B}

A distribution $T$ is a linear continuous functional defined in a space of rapidly 
decreasing test functions $\left\{ \varphi \right\} $, the so-called Schwartz space ($\mathcal{J}$),%
\begin{equation}
T:\varphi \rightarrow \left\langle T,\varphi \right\rangle \in \mathbb{C} .
\end{equation}%
In the space $\mathcal{J}$ is possible to define the Fourier transform
of a distribution. Then, the Fourier transformed distribution $\hat{T}$ is
formally defined 
\begin{equation}
\hat{T}:\varphi \rightarrow \left\langle \hat{T},\varphi \right\rangle
=\left\langle T,\hat{\varphi}\right\rangle ,
\end{equation}%
where $\hat{\varphi}$ is the Fourier transformed test function. For
instance, we can find that the Fourier transform of the one dimensional
 $\delta $-Dirac distribution is given by: $\hat{\delta}\left( k\right)
=\left( 2\pi \right) ^{-1/2}$. Moreover, since the distributional Fourier
 transform satisfies the same properties of the usual function,
  in practice, we can write $\hat{T}$, defined in $4n$-dimension, as follows
\begin{equation}
\mathcal{F}\left[ T\left( x\right) \right] \left( p\right) =\hat{F}%
\left(p\right) =\left( 2\pi \right) ^{-2n}\int
\prod\limits_{j=1}^{n}dx_{j}T\left( x\right) \exp \left[i
\sum_{l=1}^{n}p_{l}.x_{l}\right].
\end{equation}%
In order to illustrate this definition, let us consider the Fourier
transform of the following distributional product
\begin{equation}
t\left( \xi _{1},\cdots ,\xi _{r-1},\eta _{1},\cdots ,\eta _{s-1},\eta
\right) =t_{1}\left( \xi _{1},\cdots ,\xi _{r-1}\right)
\prod\limits_{j=1}^{\kappa }D_{a_{r_{j}}b_{s_{j}}}^{\left( +\right) }\left(
\xi _{r_{j}}-\eta _{s_{j}} +\eta \right) t_{2}\left( \eta _{1},\cdots ,\eta
_{s-1}\right) .
\end{equation}%
First, we define the conjugated variables so that $\xi _{i}\rightarrow p_{i}$, $\eta
_{k}\rightarrow q_{k}$, and $\eta \rightarrow q$, then the Fourier
transform is given by%
\begin{align}
\hat{t}\left( p_{1},\cdots ,p_{r-1},q_{1},\cdots ,q_{s-1},q\right) 
&=\left( 2\pi \right) ^{-\frac{4\left( r+s-1\right) }{2}}\int d^{r-1}\xi
d^{s-1}\eta d\eta \exp \left(i \sum_{k=1}^{r-1}p_{k}\xi
_{k}+i\sum_{k=1}^{s-1}q_{k}\eta _{k}+q\eta \right)  \notag \\
&\times t_{1}\left( \xi _{1},\cdots ,\xi _{r-1}\right)
\prod\limits_{j=1}^{\kappa }D_{a_{r_{j}}b_{s_{j}}}^{\left( +\right) }\left(
\xi _{r_{j}}-\eta _{s_{j}}+\eta \right) t_{2}\left( \eta _{1},\cdots ,\eta
_{s-1}\right) .  \label{eqC5}
\end{align}
Besides, we have that the Fourier transform of the distribution $%
D_{a_{r_{j}}b_{s_{j}}}^{\left( +\right) }$ is defined as%
\begin{equation}
D_{a_{r_{j}}b_{s_{j}}}^{\left( +\right) }\left( \xi _{r_{j}}-\eta
_{s_{j}}+\eta \right) =\left( 2\pi \right) ^{-2}\int dk_{j}\hat{D}%
_{a_{r_{j}}b_{s_{j}}}^{\left( +\right) }\left( k_{j}\right) \exp
\left[-ik_{j}\left( \xi _{r_{j}}-\eta _{s_{j}}+\eta \right) \right],
\end{equation}%
also by considering the identity%
\begin{align}
\sum_{i=1}^{r-1}p_{i}\xi _{i}+\sum_{k=1}^{s-1}q_{k}\eta _{k}+q\eta
&=\sum_{i\neq r_{j}}p_{i}\xi _{i}+\sum_{k\neq s_{j}}q_{k}\eta _{k}
+\sum_{j}\left( p_{r_{j}}-k_{j}\right) \xi _{r_{j}}+\sum_{j}\left( q_{\nu
_{j}}+k_{j}\right) \eta _{s_{j}}  \notag \\
&\quad +\sum_{j}k_{j}\left( \xi _{r_{j}}-\eta _{s_{j}}+\eta \right) +\left(
q-\sum_{j}k_{j}\right) \eta ,
\end{align}
we can rewrite $\hat{t}$, Eq.\eqref{eqC5}, in the following form%
\begin{align}
&\hat{t}\left( p_{1},\cdots ,p_{r-1},q_{1},\cdots ,q_{s-1},q\right) =  \notag
\\
&=\left( 2\pi \right) ^{-\frac{4\left( r-1\right) }{2}}\int d^{r-1}\xi \exp
i\left( \sum_{i\neq r_{j}}p_{i}\xi _{i}+\sum_{j}\left(
p_{r_{j}}-k_{j}\right) \xi _{r_{j}}\right) t_{1}\left( \xi _{1},\cdots ,\xi
_{r-1}\right)  \notag \\
&\times \left( 2\pi \right) ^{-\frac{4\left( s-1\right) }{2}}\int
d^{s-1}\eta \exp i\left( \sum_{k\neq s_{j}}q_{k}\eta _{k}+\sum_{j}\left(
q_{s_{j}}+k_{j}\right) \eta _{\nu _{j}}\right) t_{2}\left( \eta _{1},\cdots
,\eta _{s-1}\right)  \notag \\
&\times \left( 2\pi \right) ^{-\frac{4l}{2}}\int \prod\limits_{j=1}^{\kappa
}dk_{j}\left( 2\pi \right) ^{-\frac{4}{2}}\int d\eta \exp i\left[ \left(
q-\sum_{j}k_{j}\right) \eta \right] \prod\limits_{j=1}^{\kappa }\hat{D}%
_{a_{r_{j}}b_{s_{j}}}^{\left( +\right) }\left( k_{j}\right) .
\end{align}
Finally, identifying the respective Fourier transform of $t_{1}$ and $%
t_{2}$, this expression can be reduced to 
\begin{align}
\hat{t}\left( p_{1},\cdots ,p_{r-1},q_{1},\cdots ,q_{s-1},q\right) &=\hat{t}
_{1}\left( \cdots ,p_{i},\cdots ,p_{r_{j}}-k_{j},\cdots \right) \hat{t}%
_{2}\left( \cdots ,q_{k},\cdots ,q_{s_{j}}+k_{j},\cdots \right)  \notag \\
&\times \left( 2\pi \right) ^{-\frac{4\kappa }{2}}\int
\prod\limits_{j=1}^{\kappa }dk_{j}\delta \left( q-\sum_{j}k_{j}\right)
\prod\limits_{j=1}^{\kappa }\hat{D}_{a_{r_{j}}b_{s_{j}}}^{\left( +\right)
}\left( k_{j}\right) .
\end{align}


\end{document}